%% file: main.tex
\documentclass[preprint,aps,prb,showpacs,superscriptaddress]{revtex4-1}
\usepackage{graphicx,epsfig}
\usepackage{epstopdf}
\usepackage{bm}

\usepackage{subcaption}            
\usepackage[export]{adjustbox}
\usepackage{amsmath}               
  \allowdisplaybreaks[1]           
\usepackage{amssymb}               
\usepackage{url}                   
\usepackage{tabularx}              
\usepackage{verbatim}              
\usepackage{footnote}              
\usepackage{float}                 
\usepackage{booktabs}              
\usepackage{lipsum}                
\usepackage{ragged2e}              
\usepackage[font=small,labelfont=bf, justification=justified, format=plain]{caption} 
\usepackage[export]{adjustbox}     
\usepackage{textgreek}
\usepackage{xcolor}

\makeatletter

\renewcommand*{\p@subsection}{}

\renewcommand*{\p@subsubsection}{}
\makeatother

\newcommand{\3}{$^3$He}
\newcommand{\4}{$^4$He}

\begin{document}

\title{The motion of tracer particles in turbulent superfluid $^4$He down to the zero-temperature limit}

\author{C. O. Goodwin}
\affiliation{Department of Physics and Astronomy, The University of Manchester, Manchester M13 9PL, UK}
\email{chris.goodwin@nottingham.ac.uk}
\affiliation{School of Mathematical Sciences, The University of Nottingham, Nottingham NG7 2RD, UK}
\author{M. J. Doyle}
\affiliation{Department of Physics and Astronomy, The University of Manchester, Manchester M13 9PL, UK}
\author{J. A. Hay}
\affiliation{Department of Physics and Astronomy, The University of Manchester, Manchester M13 9PL, UK}
\author{I. Skachko}
\affiliation{Department of Physics and Astronomy, The University of Manchester, Manchester M13 9PL, UK}
\author{W. Guo}
\affiliation{Mechanical Engineering Department, Florida State University, Tallahassee, FL 32310, USA}
\affiliation{National High Magnetic Field Laboratory, Tallahassee, FL 32310, USA}
\author{P. M. Walmsley}
\affiliation{Department of Physics and Astronomy, The University of Manchester, Manchester M13 9PL, UK}
\author{A. I. Golov}
\affiliation{Department of Physics and Astronomy, The University of Manchester, Manchester M13 9PL, UK}
\email{andrei.golov@manchester.ac.uk}

\date{\today}

\begin{abstract}
An injection system for polymer particles, with diameters ranging from 1 to 6 \textmu m, has been developed for visualizing flows in superfluid \4 at temperatures down to  0.14\,K.  Using an ultrasound transducer, bursts of particles were launched into a sample of superfluid and allowed to descend under gravity.  The particles were imaged using their fluorescence in the presence of a sheet of laser light.  We report on the statistical behavior of particles during their descent, including descriptions of a mixture of smooth and erratic trajectories, indicative of the interactions with thermal excitations and quantized vortex lines.  Temperature-dependent velocity distributions were measured and analyzed, yielding Gaussian distributions with power law tails persisting into the zero temperature limit.  When sampled over increasing length scales, these distributions bifurcated into exponential for the smallest particles and bimodal Gaussian for the largest.  We also report observations of long-lived suspensions of small particles at temperatures near 1\,K, which appear to be associated with the trapping of large numbers of particles in a turbulent vortex tangle.  A method was developed for identifying and quantifying the numbers of particles bound to vortex lines, allowing for a description of the temporal dynamics of their population by an analytical model.

\end{abstract}

\maketitle

\newpage
\input{Introduction}
\input{Use_of_tracer_particles_in_He-II}
\input{Experimental_techniques}

\input{Particle_trajectories_and_velocity_statistics}
\input{Identifying_particles_bound_to_vortex_lines}

\input{Time_dependent_behaviour}
\input{Conclusions}

\input{ref}
\end{document}

%% file: Introduction.tex
\section{Introduction}
Turbulence remains a challenging problem in physics.  In classical, viscous fluids, turbulence is described as an ensemble of continuously deformable vortices of varying size and circulation \cite{pullin1998vortex}.  The continuity of the parameters describing classical vorticity presents a computational hurdle to simulating turbulent flows, making turbulence a persistently stubborn problem \cite{spalart2000strategies}.  In superfluids, including Bose-Einstein condensates and helium-II, vortex strength is quantized, delineating quantum turbulence from its classical counterpart.  Helium-II is the phase of liquid helium at temperatures below the lambda point at 2.2 K, which can be modeled as a combination of a macroscopically coherent superfluid component, which dominates in the low temperature limit, and a normal component which carries all the fluid's entropy and viscosity \cite{landau1941two, landau1941theory}.  In the superfluid component, vortices take the form of one-dimensional topological defects in the macroscopic wavefunction describing the superfluid component\cite{feynman1955}.  The vortices have a $\sim 1$\,\AA\ radius core and fixed velocity circulation $\kappa = h/m = 1.00\times10^{-3}$\,cm$^2$s$^{-1}$. Their individual dynamics can be described by the vortex filament model (VFM) \cite{schwarz1988}, and their collective behavior, when in a dense dynamic tangle, is chaotic and complex, and is dominated by intervortex interactions, such as reconnections \cite{barenghi2014introduction}.

The normal component of helium-II  consists of thermal quasiparticle excitations (phonons and rotons), moving ballistically over macroscopic distances in the low temperature limit, and with a decreasing mean free path as their number density increases with increasing temperature.  At $T \gtrsim 0.8$\,K the normal component takes the form of a viscous fluid with hydrodynamic properties which persist up to the lambda point \cite{jager1995}.  Interactions between the two fluid components are mediated by mutual friction \cite{hall1956, hall1957}, resulting from the scattering of thermal excitations from the cores of vortex filaments. In the virtual absence of mutual friction at $T \lesssim 0.4$\,K, helical perturbations of the shape of vortex lines (Kelvin waves) become omnipresent down to very small wavelengths.

The regularity of vortices in superfluids restricts the parameter space of a turbulent flow created by a vortex tangle, somehow easing the primary hurdle of simulating quantum turbulence compared to its classical counterpart.  However, the discontinuous interactions of reconnecting vortex lines introduce a rich landscape of dynamics such as Kelvin wave cascades, which are not easily described by analytical or numerical models.  For this reason, the challenge of experimentally visualizing quantized vortices to observe their evolution is an area of significant interest, particularly in the zero-temperature limit where the viscous damping effects of the normal fluid component can be neglected, and the rich physics of vortex dynamics is fully manifest.

For this work, we have successfully developed a particle injection system which allows for the dynamics of quantum turbulence to be studied at temperatures down to 0.14\,K, where the effects of mutual friction are minimal due to the rarefied normal component.  In Sec. \ref{Experimental techniques} we will detail the techniques and apparatus used to realize this experiment, including the particle injection system, imaging techniques and particle tracking methods.  In Sec. \ref{Particle trajectories and velocity statistics}, we will detail the observations made of the collective motions of large numbers of particles acquired in three distinct experimental regimes.  We will examine the distributions of particle velocities in select cases, including their dependence on the length scales over which they are measured.  In Sec. \ref{Identifying particles bound to vortex lines} we will present a method for differentiating particles bound on vortex lines in the absence of a strong counterflow of the normal and superfluid components, and in Sec. \ref{Time dependent behaviour} we will use this approach to track the number of bound particles as a function of time and temperature. 

%% file: Use_of_tracer_particles_in_He-II.tex
\section{Use of tracer particles in He-II}
The earliest use of tracer particles for flow visualization in helium-II involved neutrally buoyant hollow glass spheres which could be imaged using laser light \cite{murakami1989flow}.  Since the early 2000's, a number of experiments \cite{zhang2004, zhang2005, bewley2006, paoletti2008PowerLaw, guo2009, zmeev2013, fonda2014, mantia2014a, mantia2014b, peretti2023, minowa2025ablation} have successfully imaged the evolution and dynamics of quantized vortices in superfluid helium using small tracer particles.  Commonly used particles are micron-sized flakes of solid hydrogen, which are formed by injecting a small amount of hydrogen gas, diluted into relatively large quantities of helium, into a volume of superfluid.  The particles are made visible using a laser sheet which scatters light from the particles in the field of view of a camera.  Particles then are tracked using a combination of particle image velocimetry (PIV) for large, bulk flows, and particle tracking velocimetry (PTV) for the case where the motions of individual particles are more relevant.

Over the course of these experiments, observations have been made of particles entrained in thermal counterflows \cite{murakami1989flow, zhang2004, zhang2005, guo2009}; interacting with vortex lines \cite{mantia2014a, mantia2014b}; and densely decorating individual vortices exhibiting reconnections, Kelvin waves and in polarized arrays in states of steady rotation \cite{bewley2006, fonda2014, peretti2023}.

There are a number of benefits to the use of hydrogen flakes as tracer particles.  The small size of the flakes minimally affects the motion of vortices, they have small terminal velocities and leave no residue in the experimental cell once warm.  A crucial limitation of the method is that the small cooling power of dilution refrigerators, normally used to achieve temperatures below 1\,K, is insufficient to overcome the unavoidable substantial thermal load on the experimental cell due to the gas injection apparatus.  Alternative tracer particles have been suggested in order to reach lower temperatures, such as electron bubbles \cite{guo2007observations} or metastable He$_2^*$ excimers, excited directly out of the liquid helium using a high voltage field emission tip \cite{guo2009, zmeev2013}.  While alternative tracer particles have shown promise as a replacement to hydrogen, until now, no experiment has achieved visualization of particle-vortex interactions at temperatures below 1\,K.

%% file: Experimental_techniques.tex
\section{Experimental techniques}
\label{Experimental techniques}
\subsection{Particle injection}
\begin{figure}[ht]
   \centering
       \begin{minipage}[h]{0.55\textwidth}        \includegraphics[width=\textwidth]{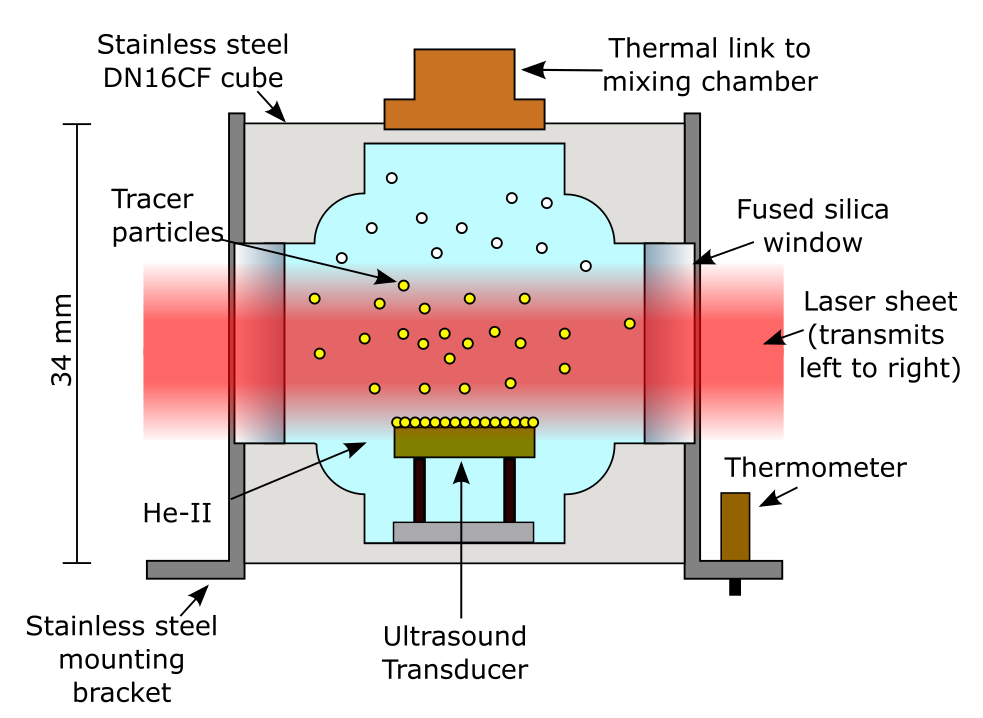}
         \end{minipage}
         \hfill
         \begin{minipage}[h]{0.44\textwidth} \includegraphics[width=\textwidth]{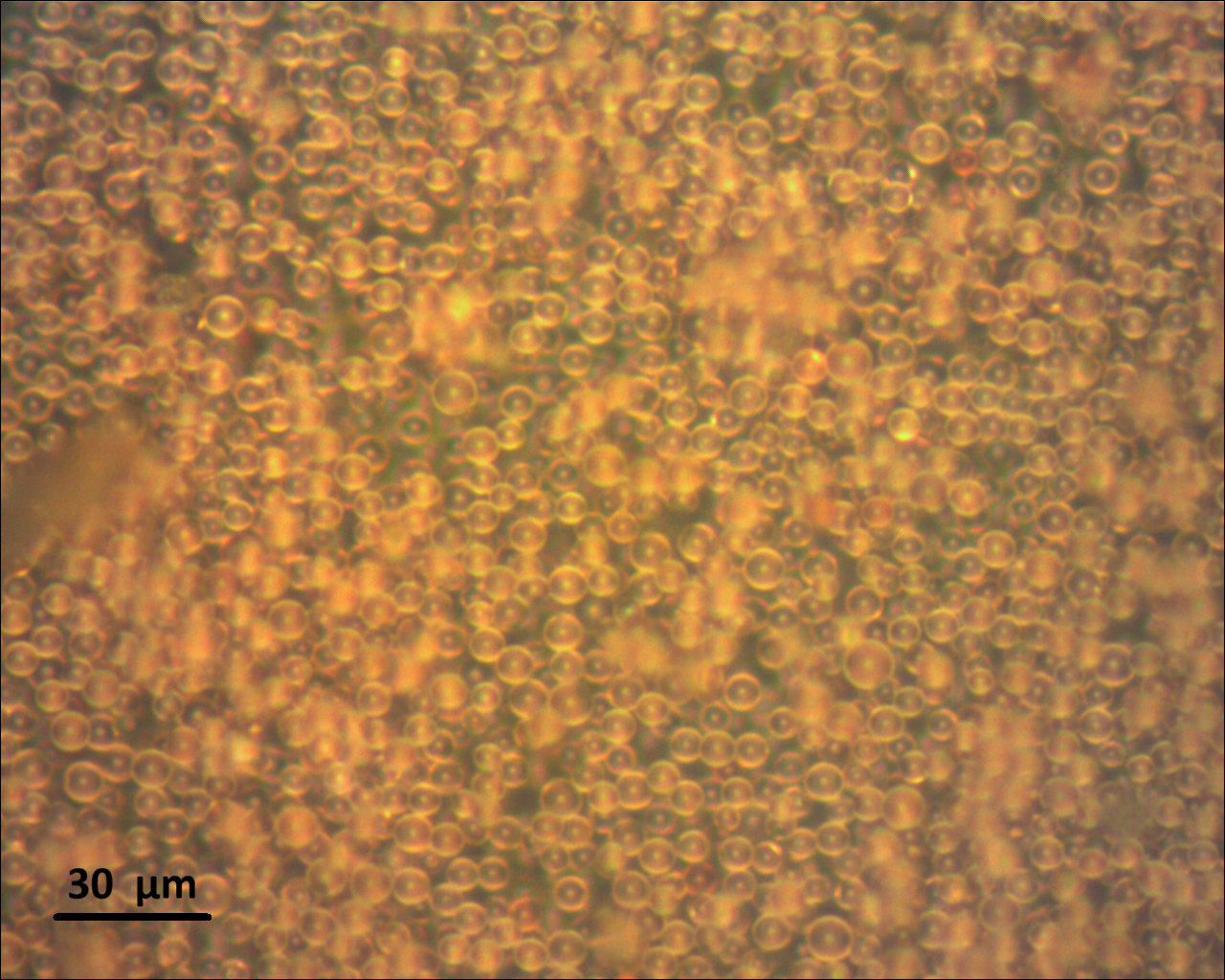}
        \end{minipage}
    \caption{Left: Schematic of the experimental cell showing the path of the laser sheet, particle injection system, positions of the cell thermometer and thermal link to the refrigerator. Right: Microscopic view of 6 \textmu m particles on the transducer surface.}
    \label{fig:experimental_cell}
\end{figure}

In our experiment, fluorescent-dyed polymer microspheres with diameters of 1--6 \textmu m were impulsively injected from the surface of an ultrasound transducer, submerged in a volume of superfluid in a completely filled experimental cell \cite{skachko2026RSI}, as shown in Fig. \ref{fig:experimental_cell}. 

The cell has a base temperature of 140\,mK while thermally anchored to the mixing chamber of a wet \3-\4 dilution refrigerator (referred to throughout as the fridge), with a minimum temperature of 12\,mK \cite{fear2013}.  The transducer, acquired from \textit{Boston Piezo Optics}, is a 12.7\,mm diameter, 1\,mm thick disk of a 36$^\circ$ Y-cut lithium niobate piezoelectric crystal with a resonant frequency of 1 MHz and a mechanical Q factor of $\sim 10^4$.  The transducer top surface is coated with many layers of dry, fluorescent microspheres at room temperature before being loaded into the cell.  The deposition of the particles onto the transducer surface is achieved using an aerosolisation technique.  A small sample of particles is loaded into a cylinder, closed at one end, and capped with the transducer, face down, at the other.  Using a small hole in the side of the cylinder, a jet of dry nitrogen gas is rapidly injected, dispersing the particles into a cloud which is captured on the transducer surface.  This technique allows particles to be deposited in approximate monolayers with minimal clumping.  Repeating this process multiple times results in a large number of particles coating the surface, sufficient for several hundred bursts of particles to be released from the transducer before significant depletion occurs.  Particles are released from the transducer by applying an AC voltage of 0.5--1.5\,kV, at the resonant frequency of the transducer, in 100\,\textmu s bursts using a gated amplifier.

Once the transducer has been loaded with particles and mounted inside the cell, it is suspended below the mixing chamber, and the fridge is cooled to base temperature.  The cell is then slowly filled with isotopically purified \4, maintaining a temperature below the lambda point.  A cylindrical capacitor, intersecting the filling line, positioned at the height of the mixing chamber plate, above the top of the cell, provides a method for judging once the cell is completely filled. 
After each firing of the transducer, a burst of tens to hundreds of particles is launched vertically into superfluid helium, so the particles can be observed within the objective's field of view illuminated by the laser sheet.  
The particles freely descend with temperature dependent terminal velocities of order 1--10\,cm\,s$^{-1}$.  At the end of an experimental run, consisting of hundreds of firings over 2--3 weeks of operation, the particles coat the interior of the cell and must be removed before reloading the transducer for the next run.  This is easily achieved using a jet of nitrogen gas or compressed air.

\subsection{Particle imaging}
\begin{figure}[ht]
    \centering    \includegraphics[width=0.75\linewidth]{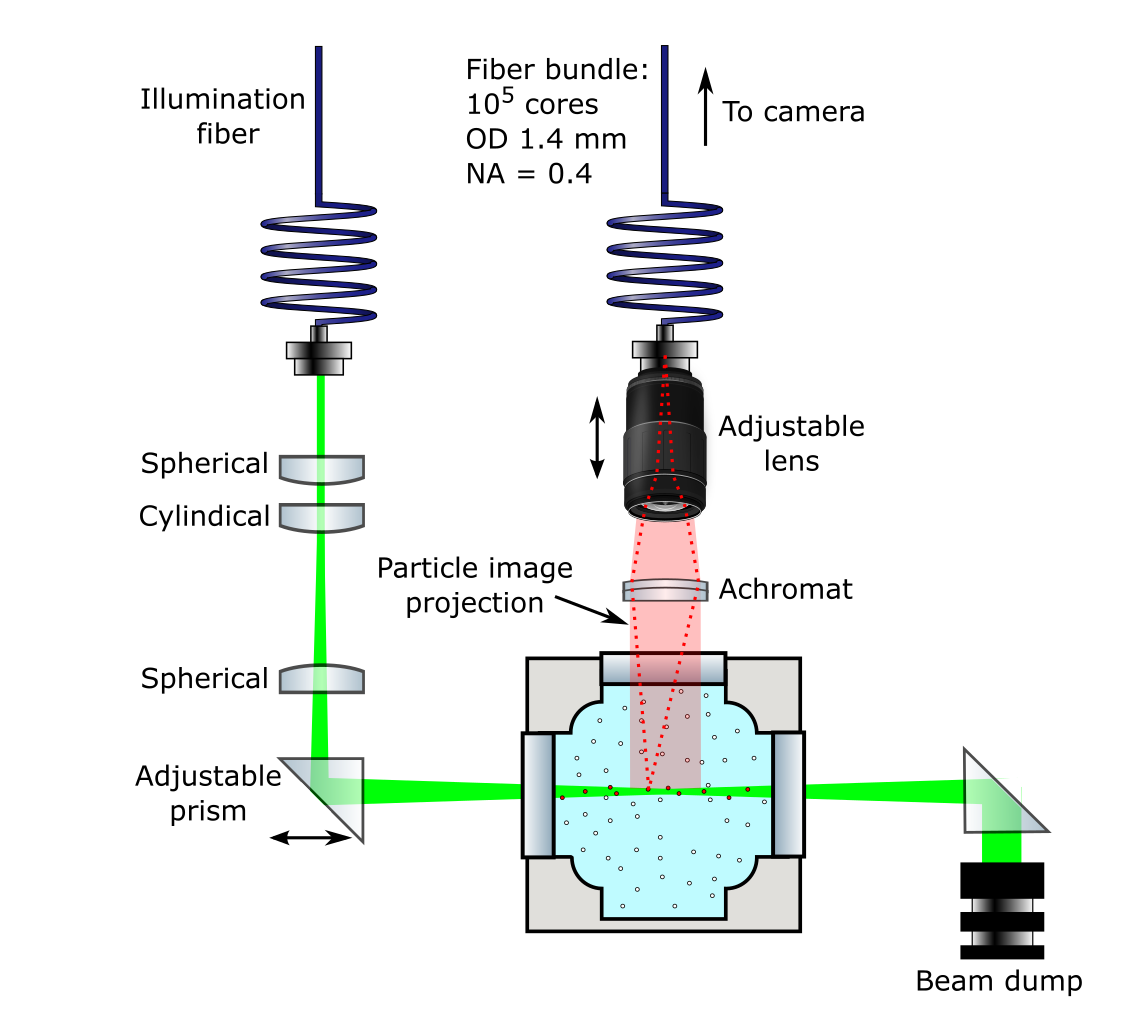}
    \caption{Schematic of the optical arrangement for admitting light to, and extracting images from the experimental cell.}
    \label{fig:optics}
\end{figure}

Particles in the cell are made to fluoresce using a 100 mW sheet of 532 nm pulsed laser light produced by a Nd:YAG laser.  The light is shaped by a combination of cylindrical and spherical lenses and diverted by a prism mounted on a remotely-controlled cryogenic piezo translational stage attached to the 4\,K stage.  The light sheet passes through anti-reflection-coated fused silica windows on opposite sides of the cell before being deposited into a beam dump, thermally anchored to a 4.2 K liquid helium bath.  The low temperature arrangement is summarized in Fig. \ref{fig:optics}, for full details see ref \cite{skachko2026RSI}.

Images of the fluorescing particles are captured with an intensified camera, located outside the cryostat, at frame rates up to 990 fps.  This requires images to be transferred out of the cryostat through a coherent fiber bundle of a 1.4\,mm diameter, containing $10^5$ parallel cores, sealed in a flexible vacuum tube. 

The laser and transducer have independent heating effects on the experimental cell.  A fast temperature spike lasting a few seconds was observed due to the firing of the transducer, followed by a long period of heating and cooling associated with the laser, lasting several minutes.  In all cases, the magnitude of the heating is on the order of tens of millikelvin, demonstrating the stability of the temperature of the cell, even in the low temperature limit.  However, it appears that the asymmetric heating from the laser on the entry and exit windows generates a horizontal thermal counterflow in the sample.   

\subsection{Particle tracking methods} \label{Particle tracking methods}
To follow the motions of particles after their dispersal in the cell, we use as our starting point a standard PTV algorithm \cite{sbalzarini2005} which identifies particles as bright spots in an image within a predefined range of intensity moments.  Particles are followed by minimizing a cost function defined as,
\begin{equation}
    \Phi = \delta x^2 +\delta y^2 + \delta m_0^2 + \delta m_2^2
    \label{cost function}
\end{equation}
where $x$ and $y$ are the image coordinates of the particle and $m_0$ and $m_2$ are the zeroth and second intensity moments respectively, and $\delta$ indicates the change between two consecutive frames.  By comparing the difference in these values for all pairs of particles across subsequent frames and finding the image pair that minimizes $\Phi$, each particle can be tracked over its trajectory through the field of view.  Once particle trajectories have been established, instantaneous velocities are measured using displacements $ \delta x$ and $\delta y$ of particle images between subsequent frames, with horizontal and vertical velocities $v_x = \delta x/\delta t$ and $v_y = \delta y/\delta t$ where $\delta t = 1.01$\,ms is the time between frames at 990\,fps.  

This approach is commonly used for PTV experiments, however the method suffers problems when particles with similar image moments cross paths, or when the inter-particle separation is small compared to the distance traveled by particles between frames.
\begin{figure}[ht]
    \centering    \includegraphics[width=0.5\linewidth]{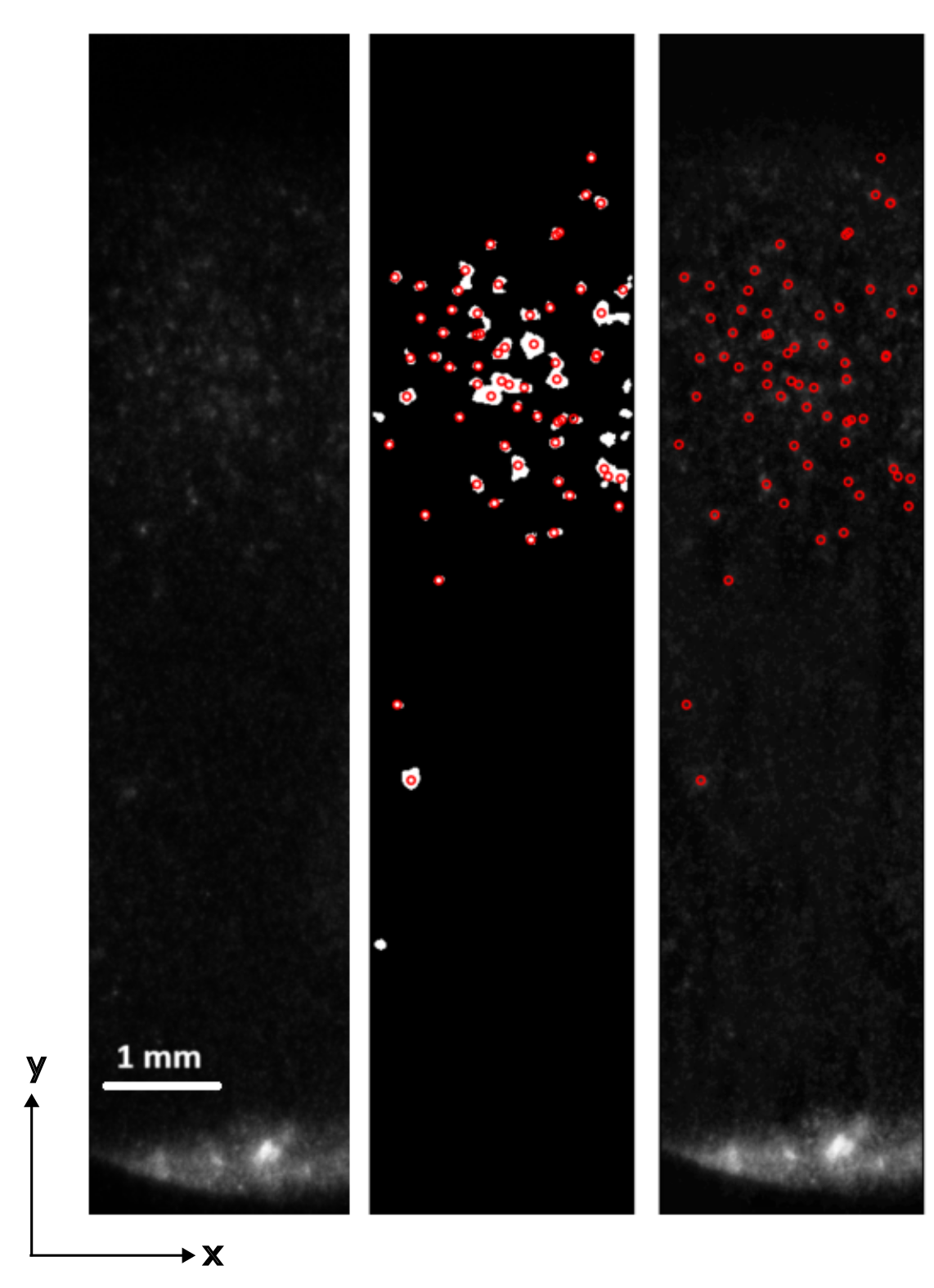}
    \caption{Masked image particle tracking example.  Left: raw image of a burst of 1-5 \textmu m particles at 1 K with transducer surface visible at the bottom of the frame. Middle: mask generated from summed frames with detections overlaid.  Right: Detections shown over raw image.}
    \label{fig:masked_detections}
\end{figure}
To overcome the limitations of the tracking algorithm an image mask was used to limit the particle search area.  The first stage to produce the mask was summing successive frames within a rolling window.  The image summing procedure produced two effects, first, the random background was flattened, greatly reducing the number of false positive signals arising from thermal electrons striking the photocathode of the camera, and also suppressing signals from particles moving perpendicular to the light sheet and those which only exist in view for a single frame.  Second, the long-lived particle images are enhanced and smeared out into short streaks, giving a visual indication of the particle trajectory.  By defining a cutoff intensity for pixels in the summed images, a time-dependent image mask is generated which restricts the search area to the tracks of long-lived particle images.  

The masking procedure significantly increased the number of true detections and suppressed the number of false positive signals. However, the method suffers the limitation that dim, fast-moving particles are suppressed with the background.  This is due to a lack of overlap in the images of such particles over successive frames. Example images of this procedure are shown in Fig. \ref{fig:masked_detections}.

%% file: Particle_trajectories_and_velocity_statistics.tex
\section{Particle trajectories and velocity statistics} \label{Particle trajectories and velocity statistics} 
Two classes of particles were used in our experiments, a sample from \textit{Thermo Fisher Scientific} with uniform 6 \textmu m diameters, and a sample of polydisperse particles from \textit{Cospheric} with diameters in the range 1--5 \textmu m and a mean of 2 \textmu m.  The mass, density and effective weight of the particles in the liquid helium background are given in Table \ref{table:particles}.
\begin{table}[ht]
    \centering
    \begin{tabular}{|l|c|c|c|c|c|c|}
        \hline
        \textbf{Particle Diameter (\textmu m)} & \textbf{1} & \textbf{2} & \textbf{3} & \textbf{4} & \textbf{5} & \textbf{6}\\
        \hline
        Density (g\,cm$^{-3}$) & 1.3 & 1.3 & 1.3 & 1.3 & 1.3 & 1.05\\
        Mass (pg) & 0.68 & 5.4 & 18 & 44 & 85 & 119\\
        Effective weight (fN) & 5.9 & 48 & 160 & 380 & 740 & 1000\\
        \hline
    \end{tabular}
    \caption{Properties of particles used during the experiment.  The effective weight gives the net downward force acting on the particles in superfluid helium accounting for buoyancy, $|\boldsymbol{F}_g + \boldsymbol{F}_b|$ where $\rho_{He} = 0.145$ g\,cm$^{-3}$.}
    \label{table:particles}
\end{table}
Across two experimental runs, three distinct particle injection regimes were investigated:
\begin{itemize}
    \item 6 \textmu m particles; 1\,s videos of small bursts containing tens of visible particles
    \item 1--5 \textmu m particles; 5\,s videos of small bursts containing tens of visible particles
    \item 1--5 \textmu m particles; 10\,s videos of large bursts containing hundreds of visible particles.
\end{itemize}
In this section, we will examine the velocity distributions 
of particles in each of the above regimes.  

\subsection{Individual particle motions} \label{Individual particle motions}
In all experiments, the particles were released with an initial vertical velocity of $\sim 70$\,cm\,s$^{-1}$ and then allowed to fall under gravity. The upward salvo of particles entrained a brief jet of the surrounding fluid, thus generating a transient large-scale turbulence of typical duration of 0.06 -- 0.1\,s, depending on temperature. Following this transient, the majority of particles moved downward with a horizontal drift, away from the warmer entry window for the illuminating light sheet. 

Several individual particles exhibited unusual motion compared to their immediate neighbors.  These stray particles came in two distinct classes:  Particles with smooth trajectories moving against the background flow of neighboring particles; and particles with jittery, or ``erratic'' trajectories with a slower drift velocity than those which appear to move freely.  At low temperatures, the erratic particles were typically observed drifting in the direction of the flow with sudden zigzagging movements, and at high temperatures, they were frequently becoming trapped in place in large numbers.
Example videos showing the various types of observed motion are provided in the supplementary material.

\subsection{Velocity statistics} \label{Velocity statistics} 
\begin{figure}[h]
    \centering    \includegraphics[width=0.75\linewidth]{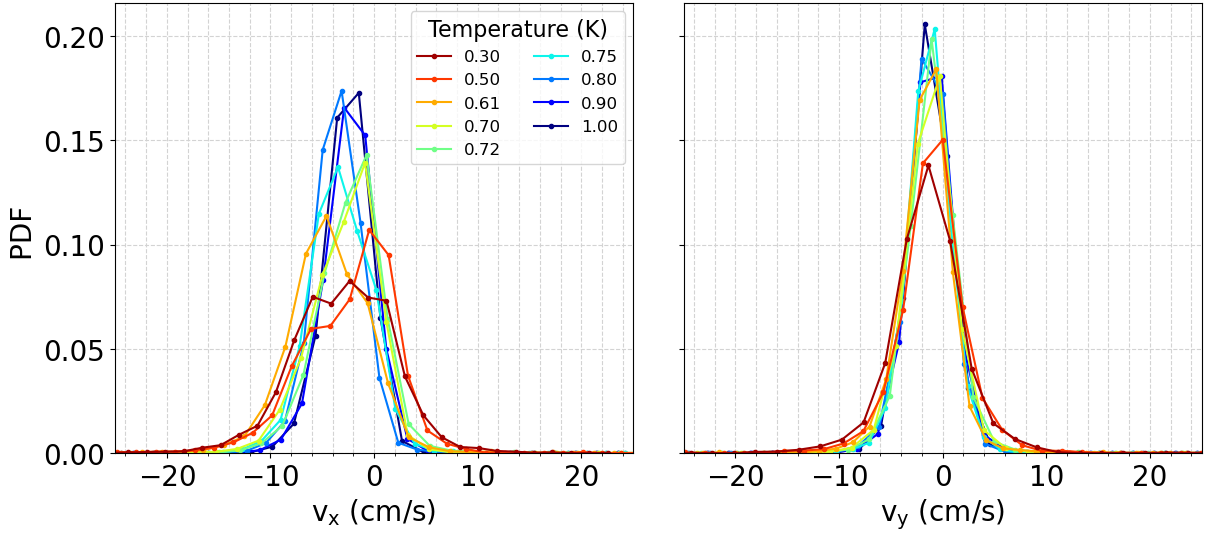}    \includegraphics[width=0.75\linewidth]{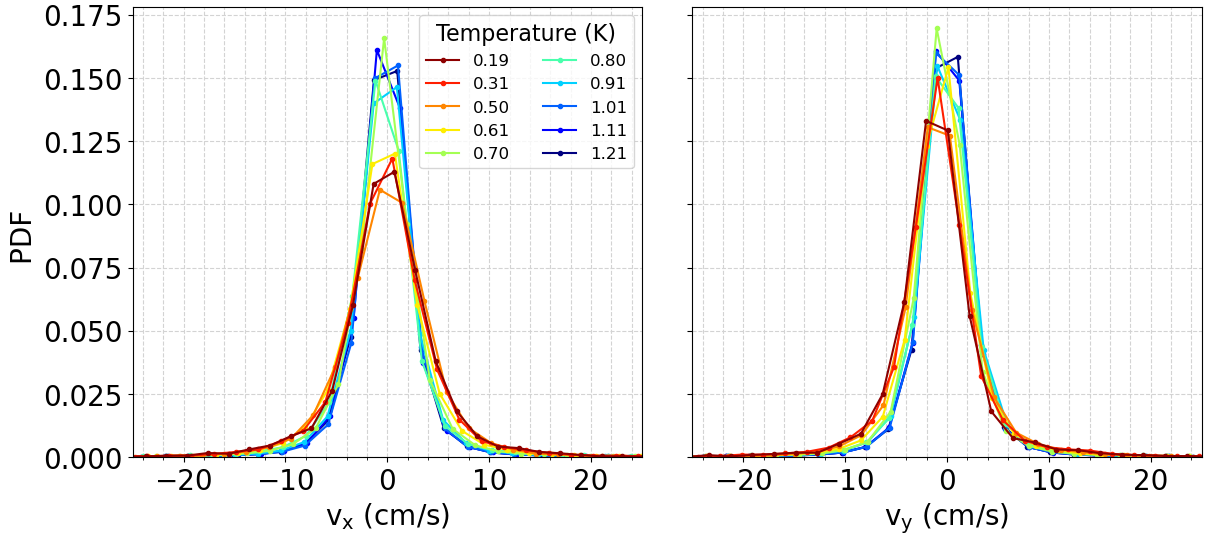}
    \caption{Velocity PDFs for 6 \textmu m particles (top) and large bursts of 1-5 \textmu m particles (bottom) for horizontal motion (left), and vertical motion (right).}
    \label{fig:PDFs}
\end{figure}

Fig. \ref{fig:PDFs} shows the area-normalized probability density functions (PDFs) derived from the velocity distributions of 6 \textmu m and 1--5 \textmu m particles, as measured using the method outlined in Sec. \ref{Particle tracking methods}, labeled by their time-averaged temperatures during observation.  The entries populating the velocity PDFs correspond to measured instantaneous velocities, per particle, per frame, after the initial turbulent transient is complete and particles have begun their descent. The tracking algorithm outlined in Sec. \ref{Particle tracking methods} suffers from poor detection efficiency for dim, fast moving particles, and does not discriminate between single particles and large clusters (however, nearly all large clusters fall down from the field of view before the turbulent transient is complete).  These limitations should be considered when asserting any qualitative interpretations of the velocity PDFs shown.

\begin{figure}[h]
    \centering    \includegraphics[width=0.8\linewidth]{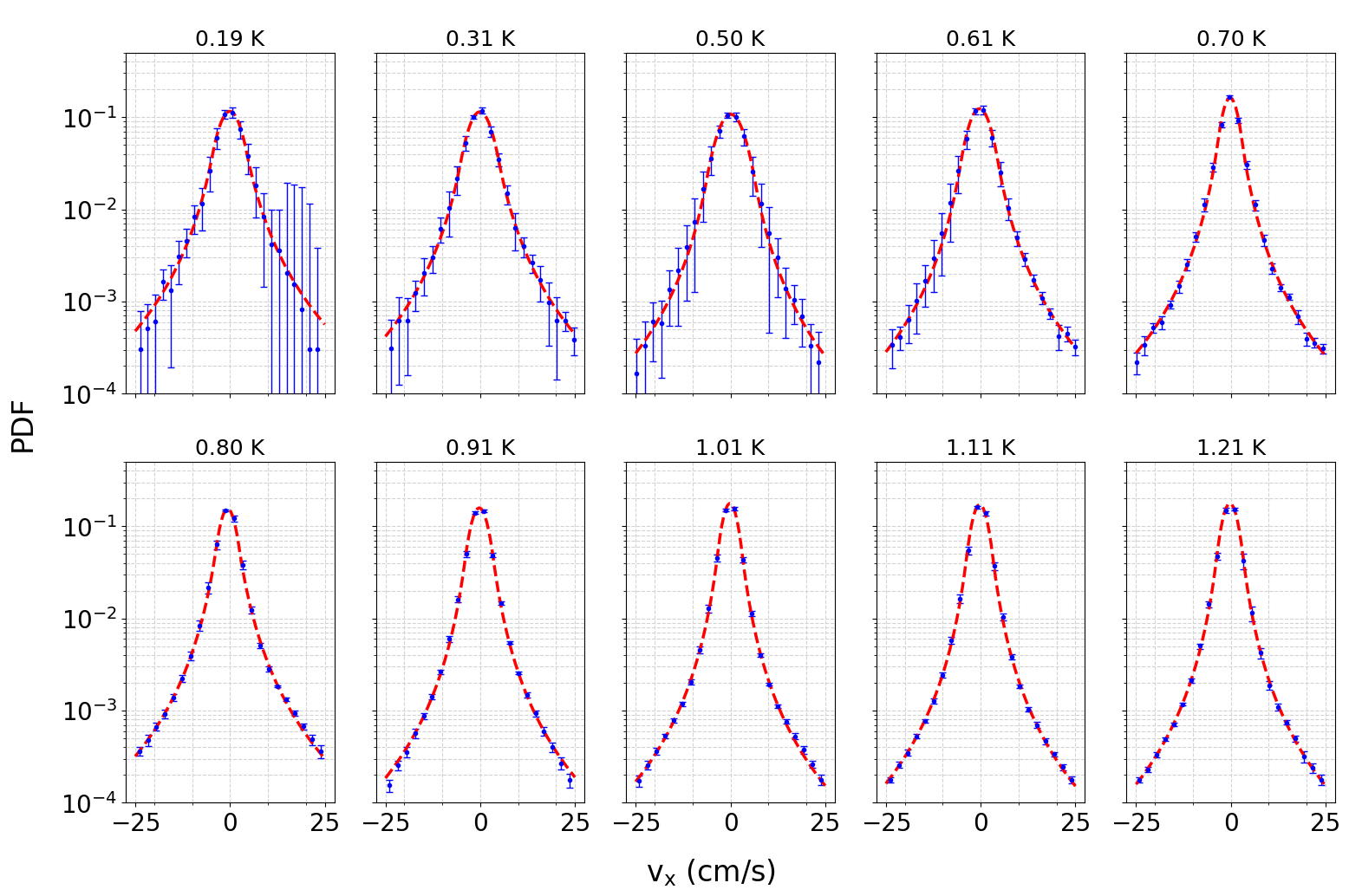}    \includegraphics[width=0.8\linewidth]{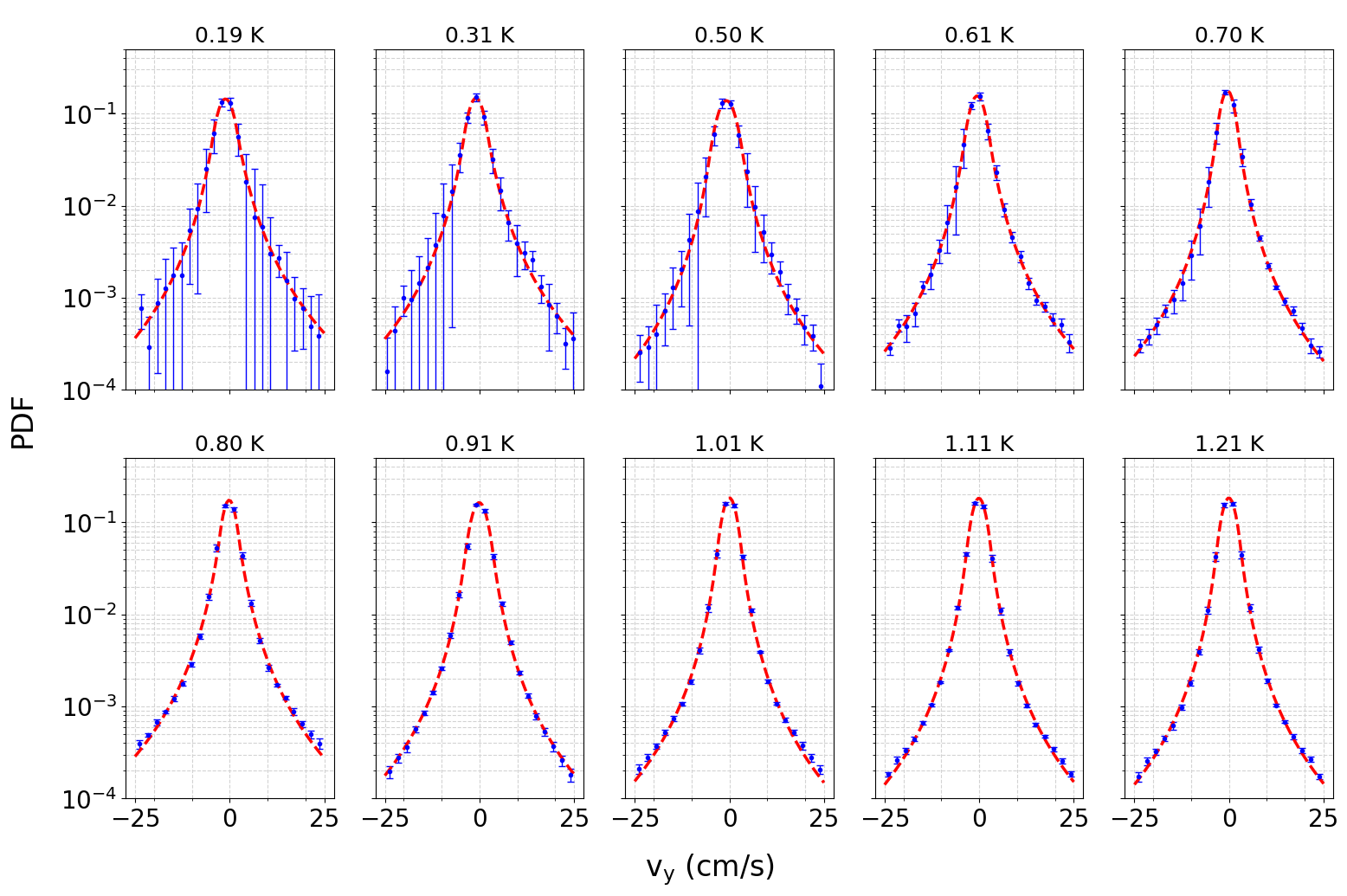}
    \caption{Horizontal (top) and vertical (bottom) velocity PDFs for large bursts of 1--5 \textmu m particles at $0.19{\rm \,K} \leq T \leq 1.2$\,K (blue points) fitted with a double-sided Crystal Ball function (red line).  Error bars show the standard error on the mean of 10 measurements at each temperature.}    \label{fig:fitted_velocities}
\end{figure}

In the case of larger (6 \textmu m) particles, temperature dependent bimodal distributions with a fast component and a slow component were often observed in the horizontal velocities at $T<0.7$\,K, suggesting particles were either free (fast particles) or trapped on vortex lines (slow particles).  Interestingly, this behavior appeared to give way to single mode, Gaussian-like distributions for smaller particles with a broader distribution of sizes. 

Despite the loss of bimodal behavior in the horizontal velocities of smaller particles, evidence remains for a combination of horizontal counterflow and particle-vortex interactions in the form of a shifted central Gaussian with power-law tails. This piecewise distribution can be parametrized using a double-sided Crystal Ball function \cite{oreglia1980study}, defined as
\begin{equation}
    f(v; N, \alpha_L, n_L, \alpha_R, n_R, \mu, \sigma) =
\begin{cases}
N A_L \left(B_L - \frac{v - \mu}{\sigma}\right)^{-n_L} & \text{for } \frac{v - \mu}{\sigma} \leq -\alpha_L \\
N e^{-\frac{1}{2} \left(\frac{v - \mu}{\sigma}\right)^2} & \text{for } -\alpha_L < \frac{v - \mu}{\sigma} < \alpha_R, \\
N A_R \left(B_R + \frac{v - \mu}{\sigma}\right)^{-n_R} & \text{for } \frac{v - \mu}{\sigma} \geq \alpha_R
\end{cases}
\label{CrystalBall}
\end{equation}
where
\begin{equation}
    A_{L/R} = \left(\frac{n_{L/R}}{|\alpha_{L/R}|}\right)^{n_{L/R}} \exp{\left(-\frac{|\alpha_{L/R}|^2} {2}\right)},\ 
    B_{L/R} = \frac{n_{L/R}}{|\alpha_{L/R}|} - |\alpha_{L/R}|.
\end{equation}
\begin{figure}[h]
    \centering    \includegraphics[width=0.9\linewidth]{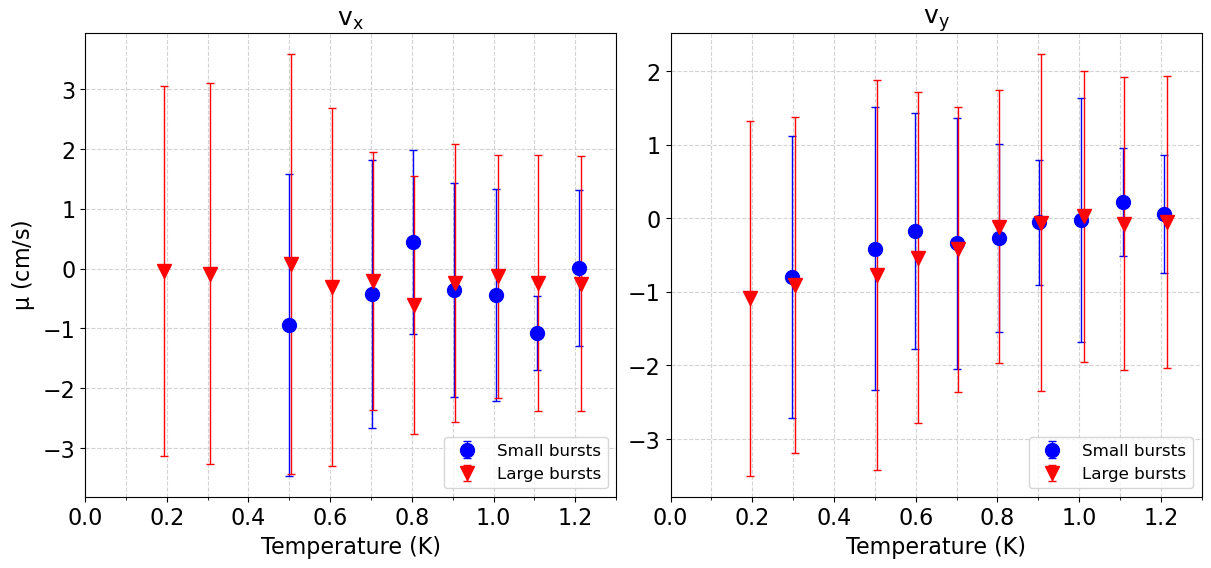}
    \caption{Mean positions for Crystal Ball function applied to 1--5 \textmu m particle velocity PDFs.  Error bars denote the standard deviation of the central Gaussian component.}    \label{fig:CrystalBallFittingParams}
\end{figure}
Here, $N$ is a normalization constant, $\mu$ and $\sigma$ are the mean and standard deviation of the Gaussian component, while $n_L$ and $n_R$ define the power law dependence of the tails.  Eqn. \ref{CrystalBall} has the benefit of being continuous and differentiable across its full domain due to the scale dependence of the power law tails on the dimensionless crossover velocities $\alpha_L$ and $\alpha_R$.  The fitted distributions for the velocities of 1--5 \textmu m particles in large bursts are shown in Fig.\,\ref{fig:fitted_velocities} as an example.  In a number of previous experiments \cite{paoletti2008PowerLaw, mantia2014a, mantia2014b, vsvanvcara2018visualization, vsvanvcara2021ubiquity, guo2010visualization, mastracci2019characterizing} distributions of this type have been observed with $n_L = n_R = 3$.  In these results, the shape of the distribution has been attributed to a combination of classical velocity fluctuations,  vortex dynamics such as reconnections and the divergence of the field velocity field near the vortex core.  The fits shown in Fig. \ref{fig:fitted_velocities} maintain the convention $n_{L/R} = 3$.  The temperature dependent parameters describing the position and width of the central Gaussian, for both $x$ and $y$ velocities of 1--5 \textmu m particles, are shown in Fig.\,\ref{fig:CrystalBallFittingParams}. 

The temperature dependence of the fitting parameters show a mean velocity ($\mu$) in $x$ and $y$ that tends toward zero with increasing temperature for 1--5 \textmu m particles.  A slight narrowing of the velocity distributions ($\sigma$) as temperature increases in all regimes suggests more coherent motion of the particle ensemble as the density of the normal component increases.  The crossover points $\alpha_{L/R}$ remain almost constant with temperature, although the different experimental regimes show small differences.  The mean values of $\alpha_{L/R}$ for small and large bursts of small particles are shown in Table \ref{table:CBAlpha}.  
\begin{table}[ht]
    \centering
    \begin{tabular}{|l|c|c|c|c|}
        \hline
        \textbf{Experiment} & $\alpha_L (v_x)$ & $\alpha_R (v_x)$ & $\alpha_L (v_y)$ & $\alpha_R (v_y)$\\
        \hline
        1--5 \textmu m (small bursts) & $1.31 \pm 0.08$ & $1.52 \pm 0.11$ & $1.68 \pm 0.10$ & $1.68 \pm 0.06$\\
        1--5 \textmu m (large bursts) & $1.37 \pm 0.03$ & $1.45 \pm 0.05$ & $1.44 \pm 0.05$ & $1.42 \pm 0.04$\\
        \hline
    \end{tabular}
    \caption{Mean values of $\alpha_{L/R}$ for particle velocities fitted by the double sided Crystal Ball function.}
    \label{table:CBAlpha}
\end{table}

\subsection{Length scale dependence} \label{Length scale dependence}
Particle velocities are sensitive to noise generated by small jumps made by particles between frames, particularly in the case of those with small velocities and erratic trajectories.  
\begin{figure}[h]
    \centering    \includegraphics[width=0.9\linewidth]{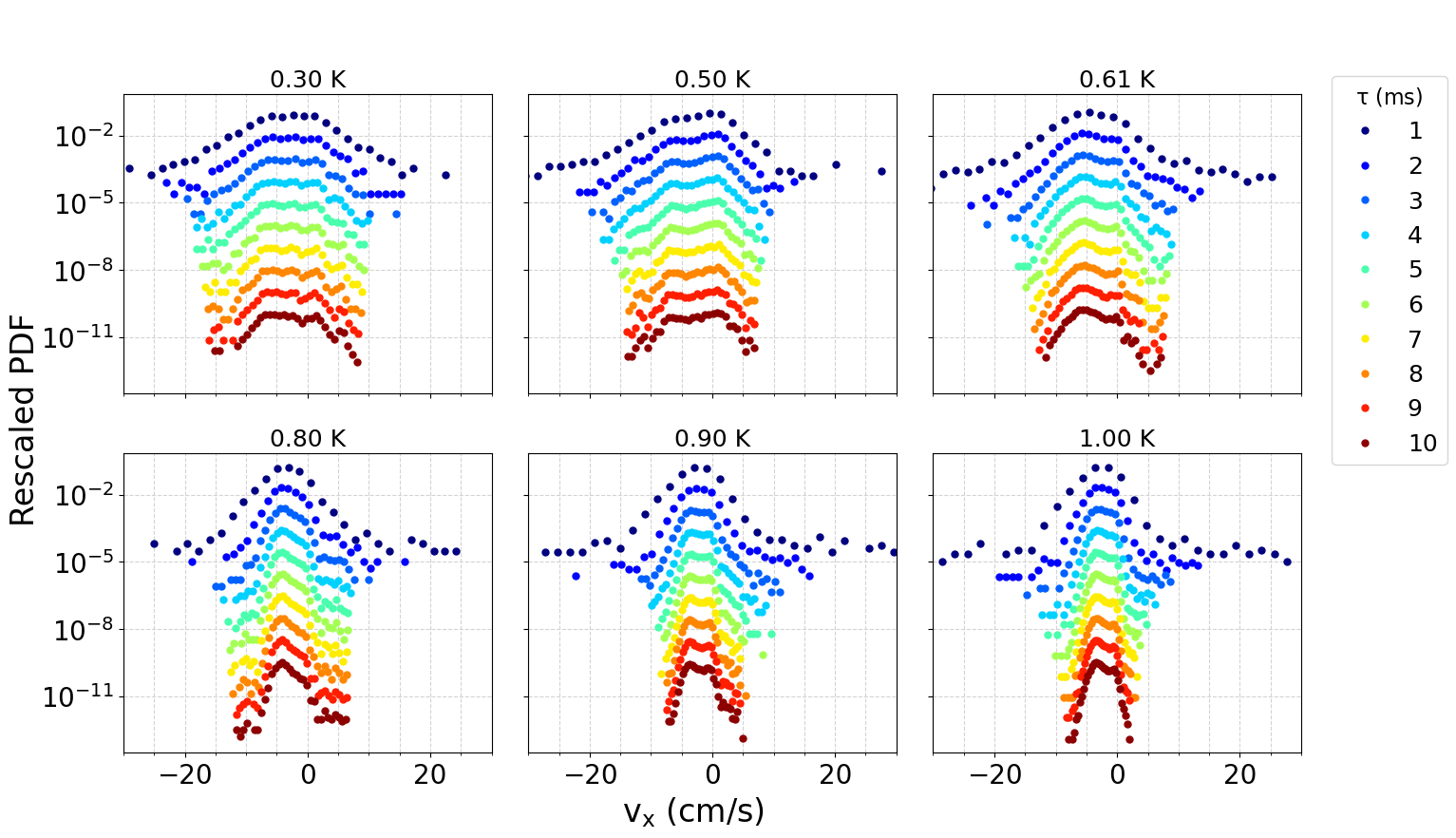}    \includegraphics[width=0.9\linewidth]{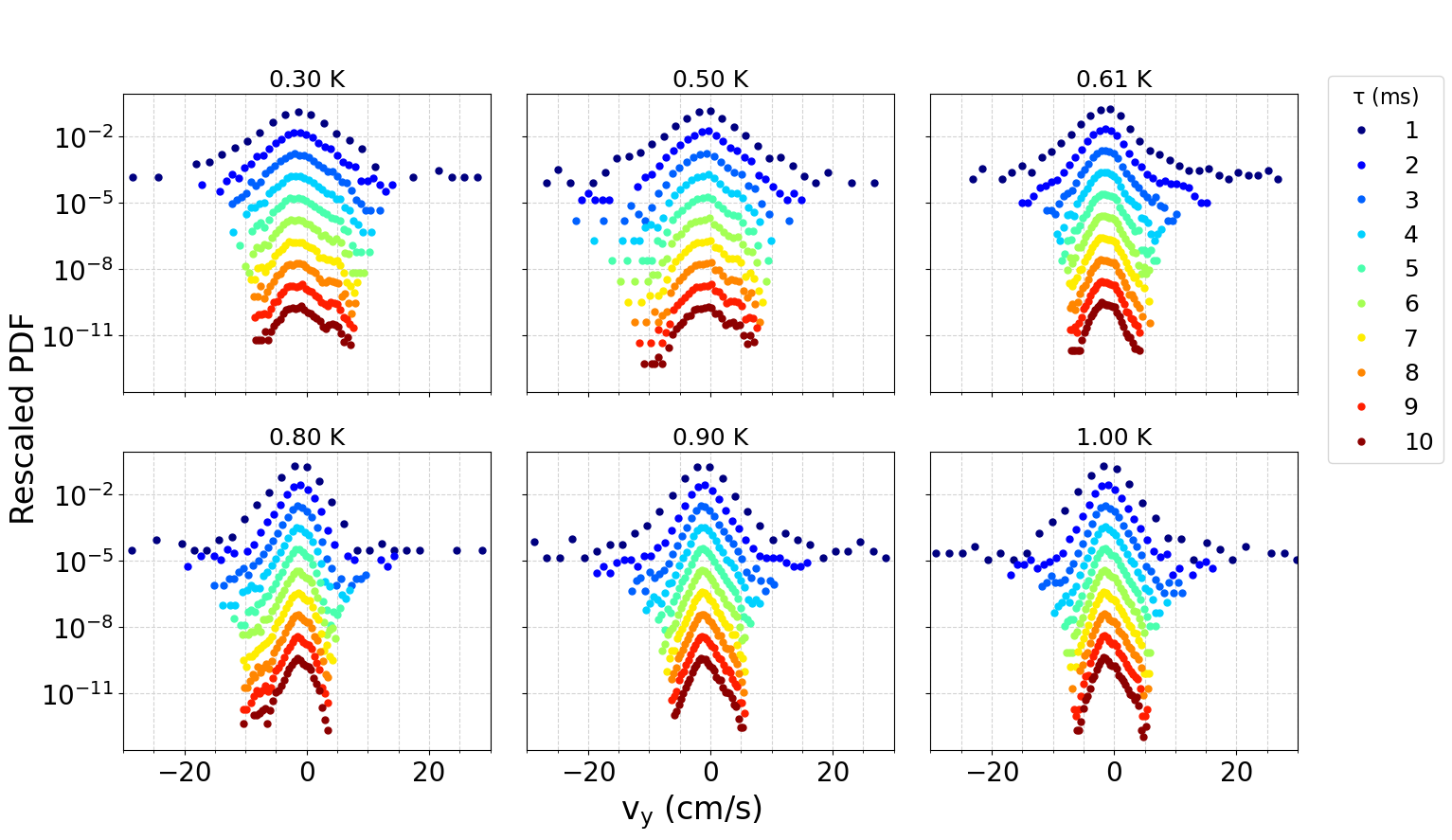}
    \caption{Horizontal (top) and vertical (bottom) velocity PDFs for 6 \textmu m particles smoothed over 1--10 frame rolling windows.  For each subsequent number of frames, the PDF is rescaled down one decade for clarity.}    \label{fig:smoothed_velocities_6um}
\end{figure}
\begin{figure}[h]
    \centering    \includegraphics[width=0.9\linewidth]{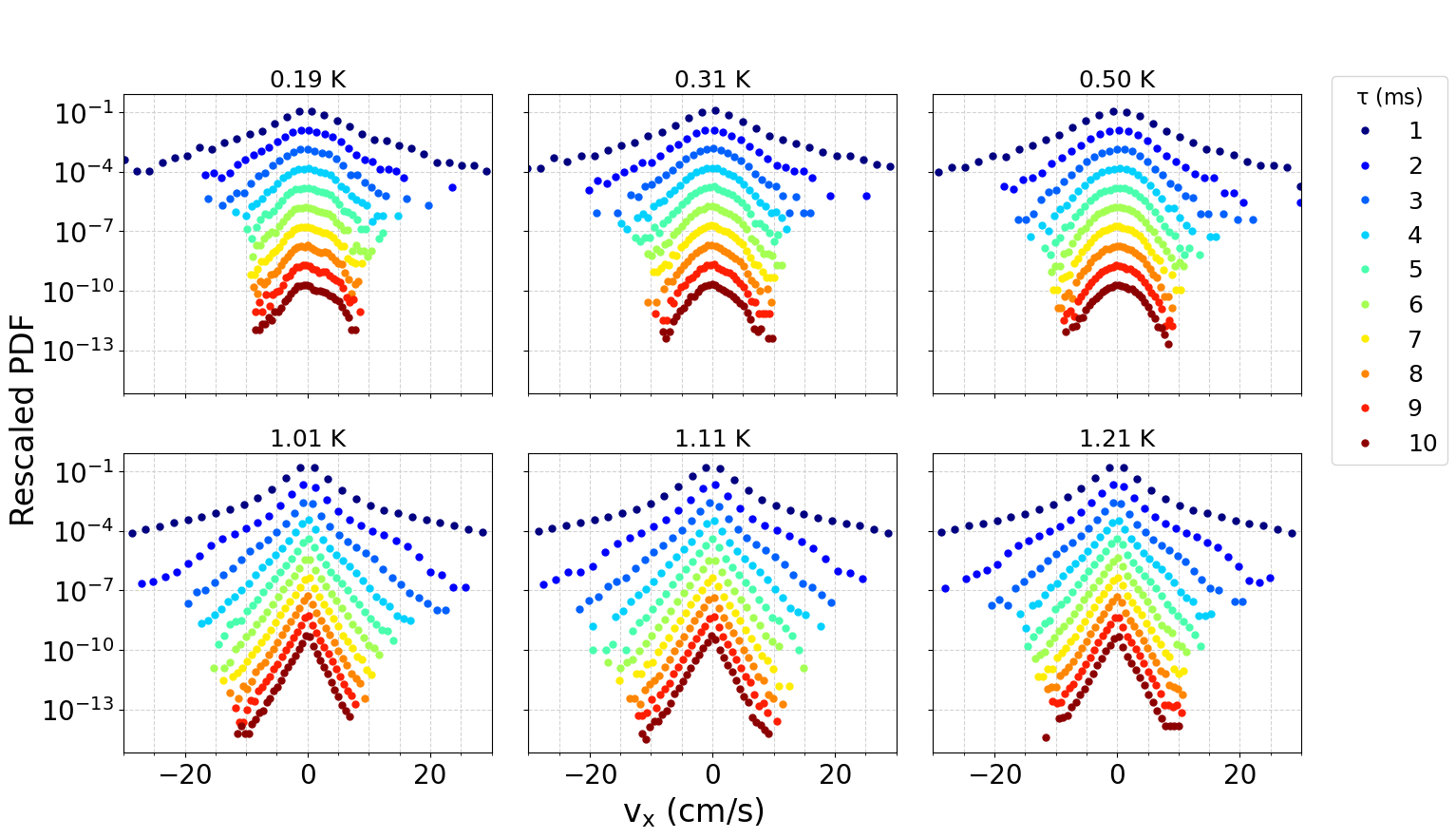}    \includegraphics[width=0.9\linewidth]{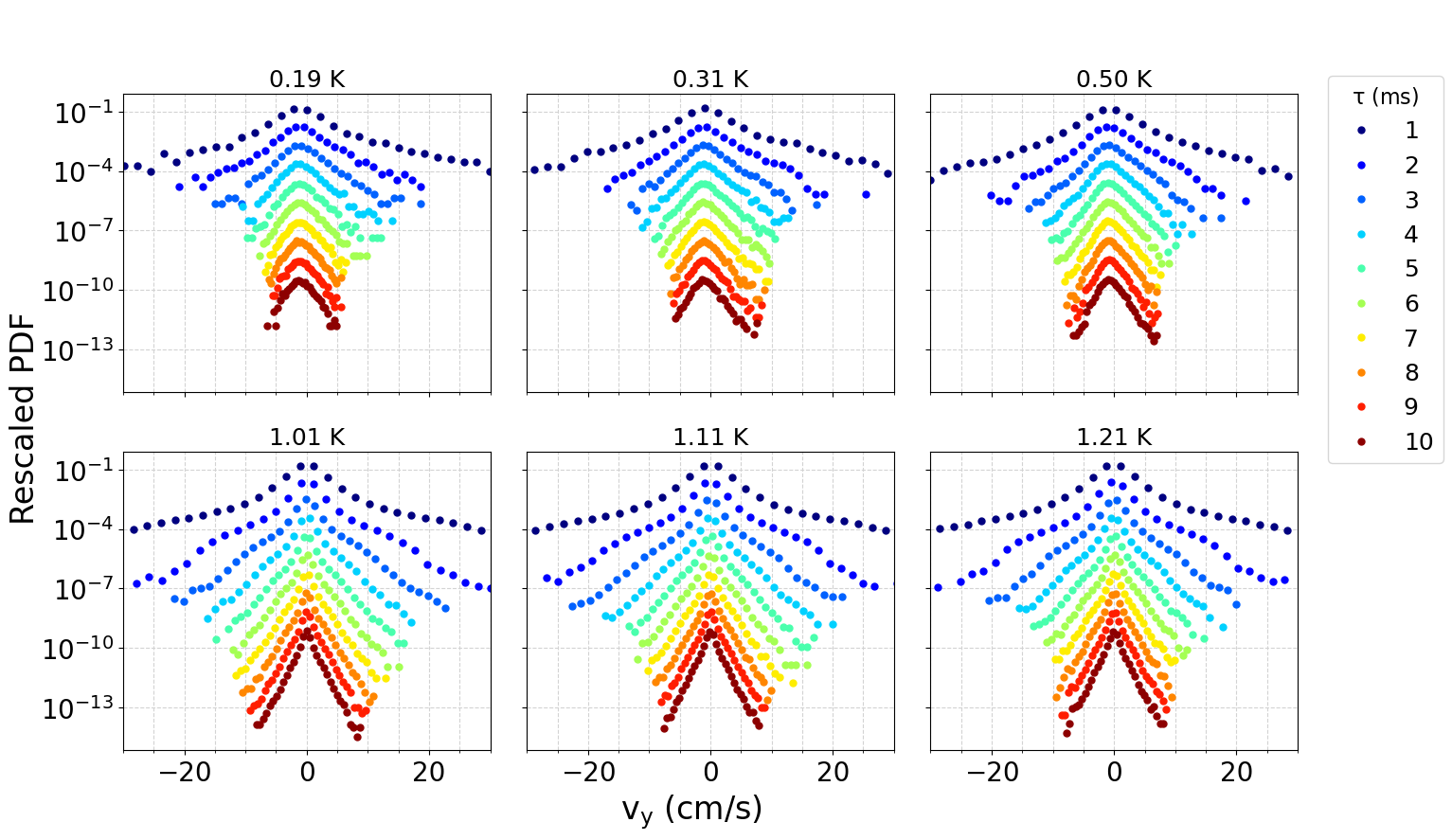}
    \caption{Horizontal (top) and vertical (bottom) velocity PDFs for 1--5 \textmu m particles smoothed over 1--10 frame rolling windows.  For each subsequent number of frames, the PDF is rescaled down one decade for clarity.}    \label{fig:smoothed_velocities_1-5um}
\end{figure}
\begin{figure}[h]
    \centering    \includegraphics[width=0.7\linewidth]{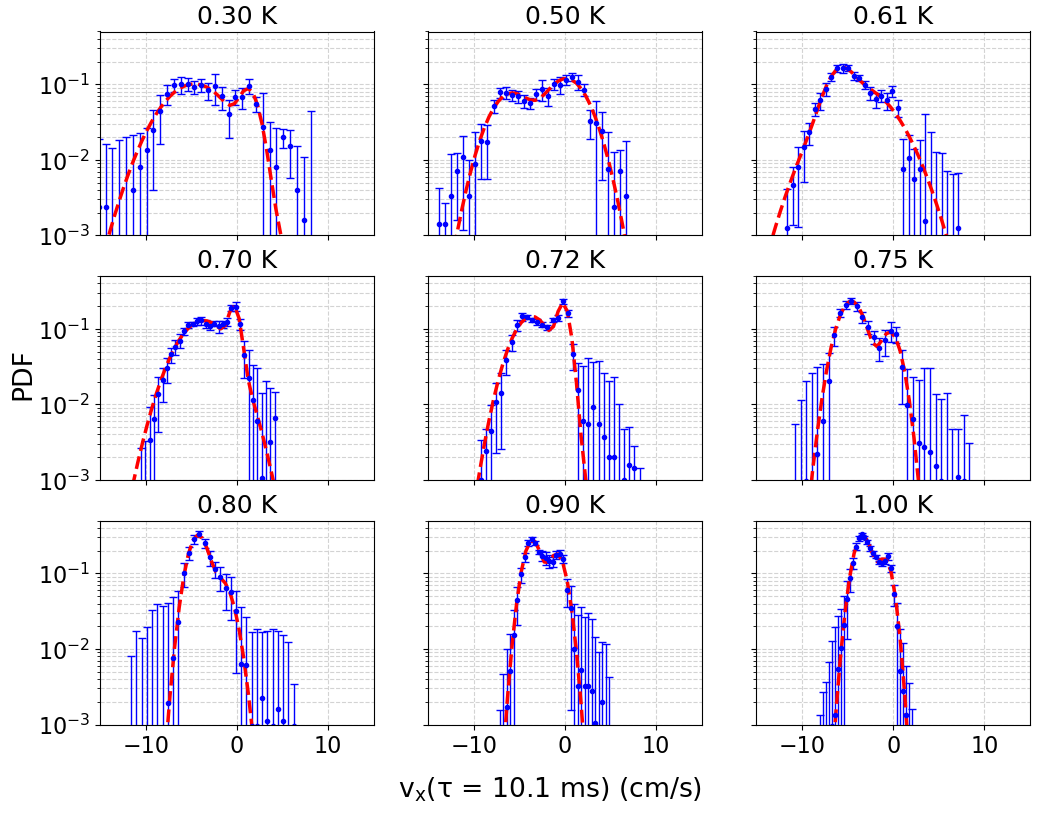}      \includegraphics[width=0.7\linewidth]{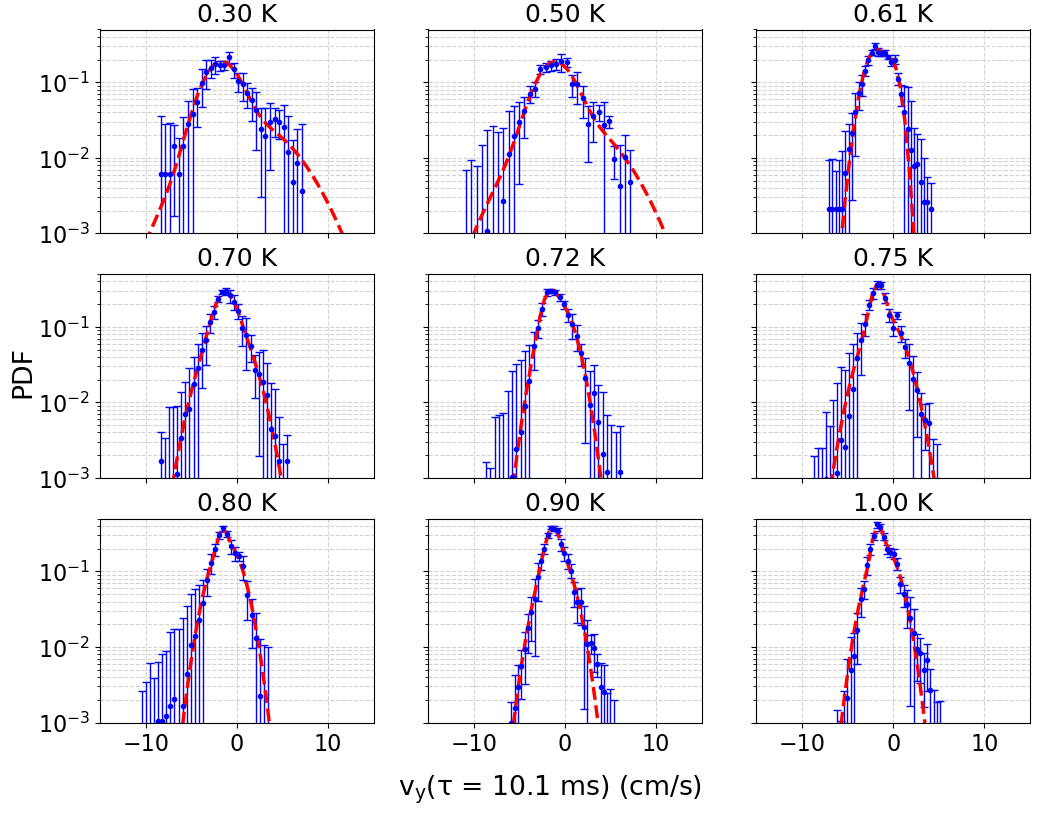}
    \caption{Fitted bimodal Gaussian particle horizontal (top) and vertical (bottom) velocity PDFs for 6 \textmu m particles with $\tau = 10.1$ ms.}  \label{fig:bimodal_gussians_6um}
\end{figure}
\begin{figure}[h]
    \centering    \includegraphics[width=0.75\linewidth]{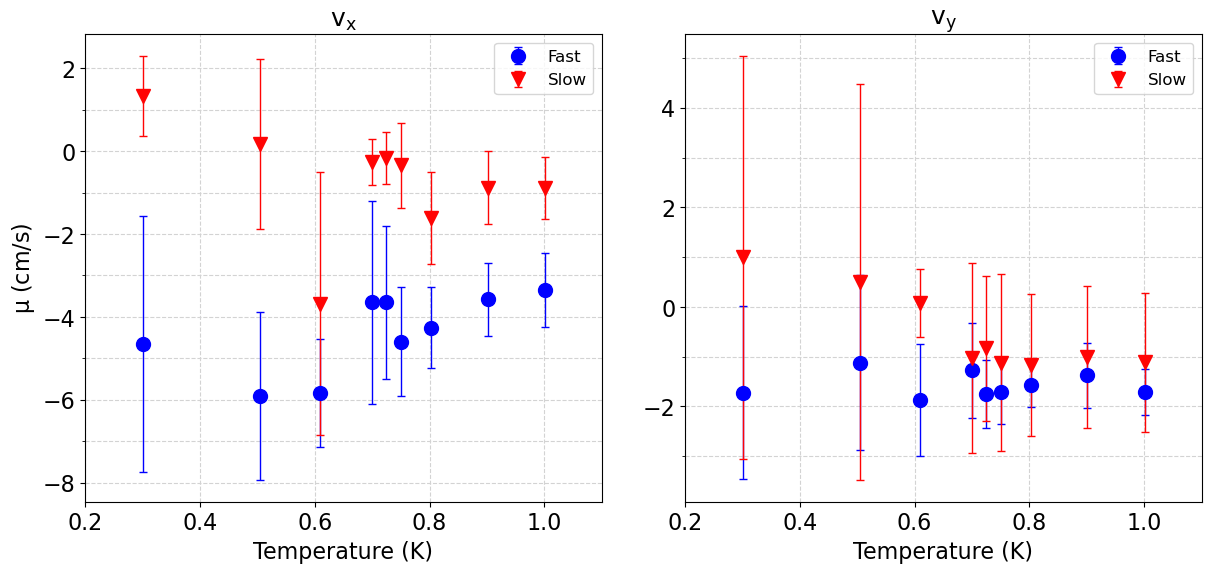}
    \caption{Bimodal Gaussian fitting parameters for PDFs of smoothed velocities of 6 \textmu m particles.  $\mu_1$ and $\mu_2$ correspond to the mean position of each Gaussian component, with error bars indicating their respective standard deviations $\sigma_1$ and $\sigma_2$.}  \label{fig:bimodal_fitting_parameters_6um}
\end{figure}
To observe overall drift velocities and better characterize the motions of free or trapped particles, a rolling window, $\tau$, is defined over which the velocities are smoothed in time, yielding $v_i(\tau) = \delta x_i/\tau$, where $\delta x_i$ is the distance moved in the $x_i$ direction in time $\tau$.  Figs.\,\ref{fig:smoothed_velocities_6um} and \ref{fig:smoothed_velocities_1-5um} show the results of this smoothing procedure for 6 \textmu m and 1--5 \textmu m particles respectively.  For each subsequent value of $\tau$ the PDF is rescaled down by one decade for easy comparison.  It was observed that as $\tau$ increases the distribution narrows and, most significantly in the case of the smaller particles at higher temperatures, the distribution tends from a Gaussian with power law tails to one with exponential tails.  The 6 \textmu m particle distributions narrow but increasingly display a bimodal shape at all temperatures, including a minor splitting into fast and slow components in the vertical velocities.

A bimodal Gaussian distribution was fitted to the smoothed velocities of the 6 \textmu m particles, revealing temperature dependent fast and slow components, particularly in the case of their horizontal motion.  This is taken as an indication of 2 populations of particles either moving freely under the influence of the normal component, or strongly interacting with vortex lines. Fig. \ref{fig:bimodal_gussians_6um} shows the fitted velocity PDFs and Fig. \ref{fig:bimodal_fitting_parameters_6um} shows the fitting parameters measured for the distributions.

In previous experimental works \cite{mantia2014a, mantia2014b} a scale dependence on velocity distributions has been observed at higher temperatures and attributed to the measurements being made on length scales exceeding the intervortex spacing, suppressing the power law tails associated with vortex interactions such as reconnections.  However in these previous cases, the resultant distributions were Gaussian, not exponential.  Interestingly, such exponential distributions have been predicted in numerical simulations as a signature of the turbulent velocity components of vortex lines as described by the vortex filament model \cite{baggaley2014acceleration, galantucci2025quantum}.

\begin{figure}[h]
    \centering    \includegraphics[width=0.85\linewidth]{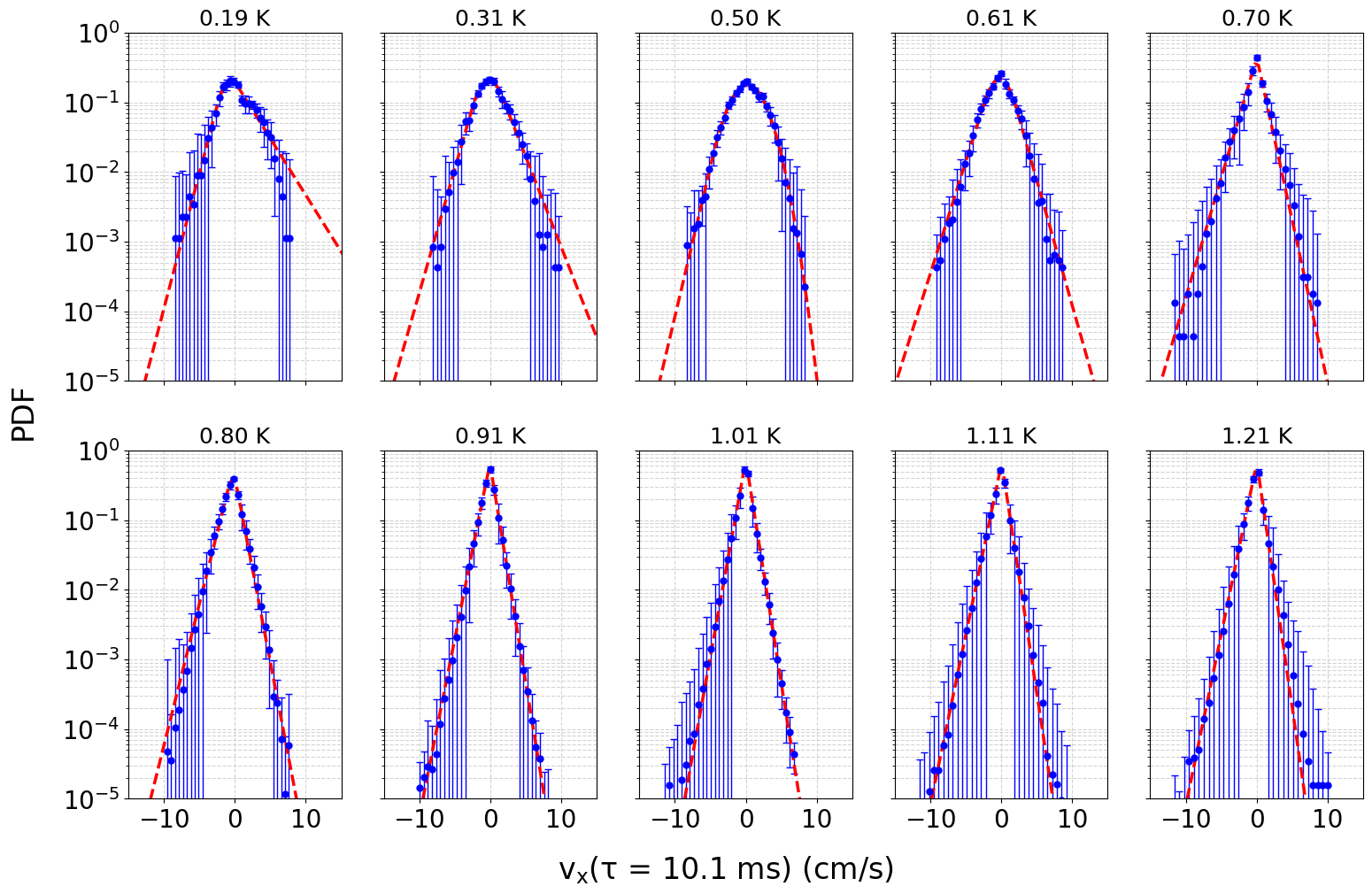}
    \includegraphics[width=0.85\linewidth]{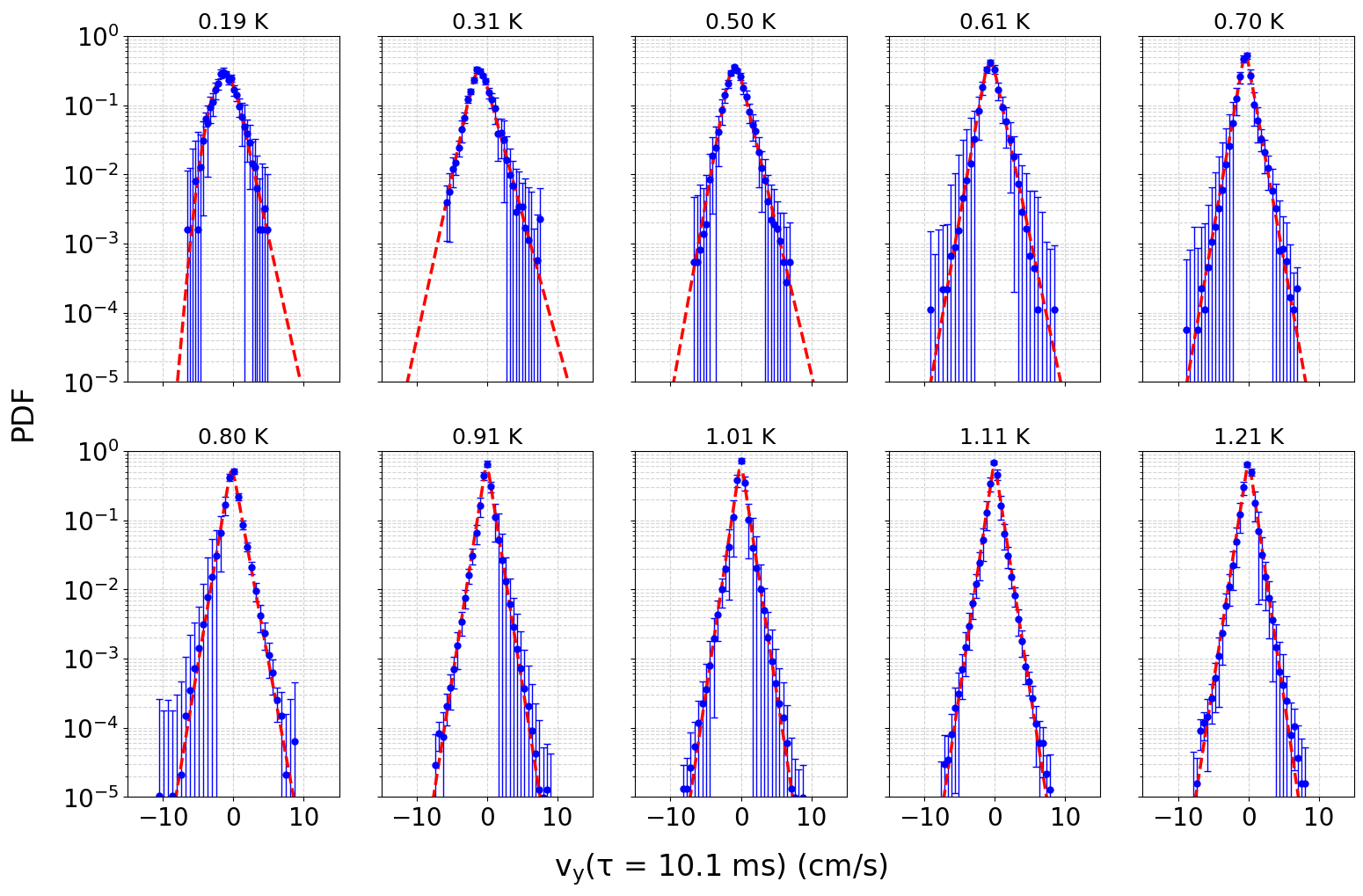}
    \caption{Horizontal (top) and vertical (bottom) smoothed velocity PDFs for large bursts of 1--5 \textmu m particles with $\tau = 10.1$\,ms, fitted with double-shouldered Gaussian exponential.}    \label{fig:double_shouldered_GuassExp_1-5um}
\end{figure}

\begin{figure}[h]
    \centering    \includegraphics[width=0.9\linewidth]{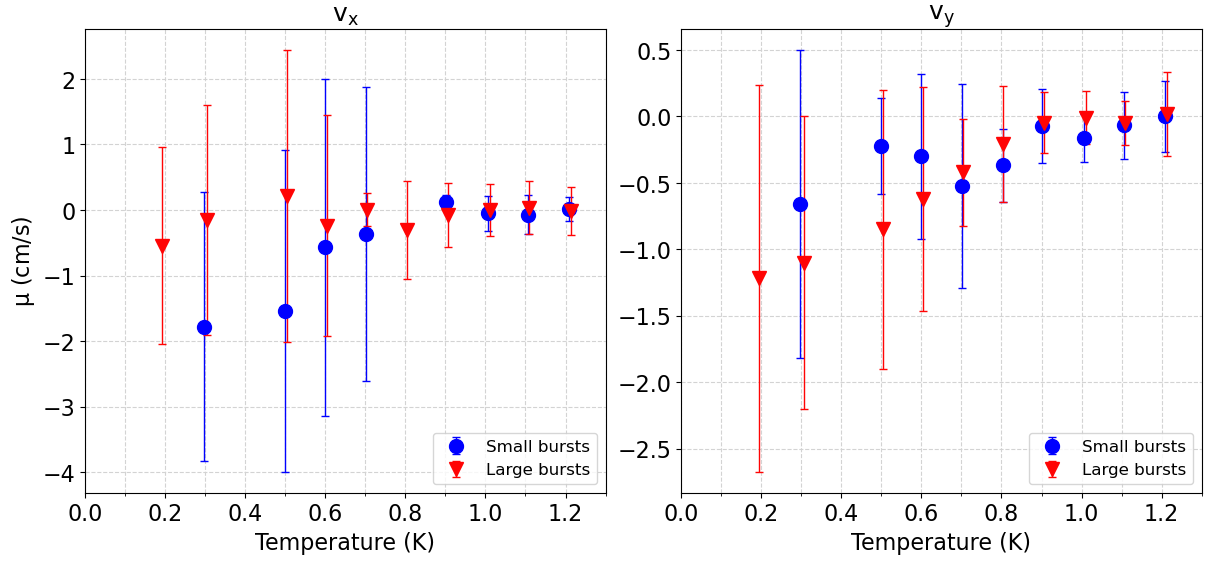}
    \caption{Mean positions for double-shouldered Gaussian exponential function applied to 1--5 \textmu m smoothed particle velocity PDFs.  Error bars denote the standard deviation of the central Gaussian component.}    \label{fig:double_shouldered_GuassExp_params}
\end{figure}

To reflect the change in the distribution, the smoothed velocities with $\tau = 10.1$\,ms were fitted using a double-shouldered Gaussian exponential \cite{das2016simple}, defined as,
\begin{equation}
    f(v; N, k_L, k_R, \mu, \sigma) =
\begin{cases}
N e^{\frac{k_L^2}{2} + k_L \frac{v - \mu}{\sigma}} & \text{for } \frac{v - \mu}{\sigma} \leq -k_L \\
N e^{-\frac{1}{2} \left(\frac{v - \mu}{\sigma}\right)^2} & \text{for } -k_L < \frac{v - \mu}{\sigma} < k_R \\
N e^{\frac{k_R^2}{2} + k_R \frac{v - \mu}{\sigma}} & \text{for } \frac{v - \mu}{\sigma} \geq k_R.
\end{cases}
\label{GuassExp}
\end{equation}
Similar to Eqn. \ref{CrystalBall}, the function is a piecewise distribution which is continuous and differentiable over its domain, with exponential tails replacing those defined by a power law.  Examples fits for the large burst velocities are shown in Fig. \ref{fig:double_shouldered_GuassExp_1-5um} with fitting parameters for all 1--5 \textmu m particles shown in Fig. \ref{fig:double_shouldered_GuassExp_params}.

As with the results shown in Fig. \ref{fig:CrystalBallFittingParams}, the smoothed velocities tend toward a mean of zero value at high temperatures, in contrast to the heavier 6 \textmu m particles which continue to descend with a mean $v_y \approx 1$\, cm\,s$^{-1}$ and $v_x \approx 2.5$\, cm\,s$^{-1}$.  And again, as the normal component density increases, the flow becomes more coherent with decreasing $\sigma$ values.  The $k_{L/R}$ values remain nearly constant with temperature, with significant differences for small and large bursts as shown in Table \ref{table:k_GE}.
\begin{table}[ht]
    \centering
    \begin{tabular}{|l|c|c|c|c|}
        \hline
        \textbf{Experiment} & $k_L (v_x)$ & $k_R (v_x)$ & $k_L (v_y)$ & $k_R (v_y)$\\
        \hline
        1--5 \textmu m (small bursts) & $1.47 \pm 0.35$ & $1.47 \pm 0.25$ & $0.98 \pm 0.14$ & $1.03 \pm 0.16$\\
        1--5 \textmu m (large bursts) & $0.77 \pm 0.19$ & $0.84 \pm 0.22$ & $1.59 \pm 0.45$ & $0.91 \pm 0.05$\\
        \hline
    \end{tabular}
    \caption{Mean values of $k_{L/R}$ for particle velocities fitted by the double shouldered Gaussian exponential function.}
    \label{table:k_GE}
\end{table}

Large error bars in the smoothed velocity PDFs and several fitting parameters for the double-shouldered Gaussian exponential distributions, are indicative of the reduced number of particles with sufficiently long tracks to measure over the increased period of $\tau = 10.1$\,ms.  A more detailed analysis and modeling of the behavior of particles in each of the measurement schemes offered will be forthcoming in additional publications.  For now, we offer these results without interpretation as a progress report on our findings.

%% file: Identifying_particles_bound_to_vortex_lines.tex
\section{Identifying particles bound to vortex lines} \label{Identifying particles bound to vortex lines}
In the case of strong thermal counterflows, the particles' velocity component parallel to the direction of flow provides a discriminatory framework for free particles from those trapped on vortex lines
\cite{mastracci2019characterizing}.  The velocity distributions of such particles typically form a bimodal distribution, with a fast component in the direction of the normal fluid flow and a slow component comprised of particles trapped on vortex lines.  In Fig.\,\ref{fig:PDFs}, hints are seen of such a distribution for the case of 6\,\textmu m particles.  
\begin{figure}[h]
    \centering    \includegraphics[width=\linewidth]{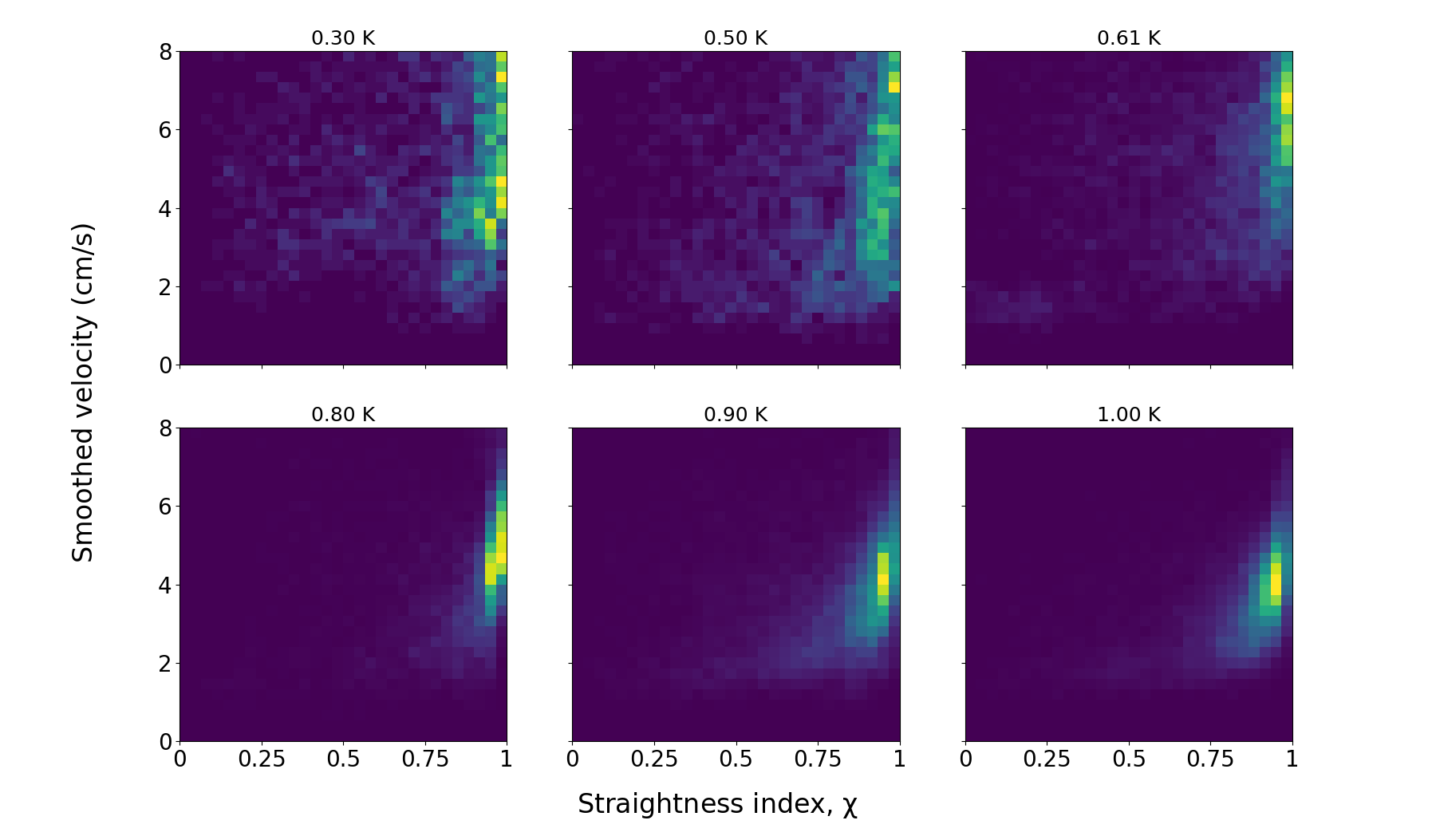}
    \caption{Particle binding parameter space shown for small bursts of 6 \textmu m particles. Smoothed velocity defined as $v = \delta r/\tau$ with $\tau = 10.1$ ms.  Heat map indicates linearly scaled occupation.}    
    \label{fig:binding space 6 um}
\end{figure}
\begin{figure}[h]
    \centering    \includegraphics[width=\linewidth]{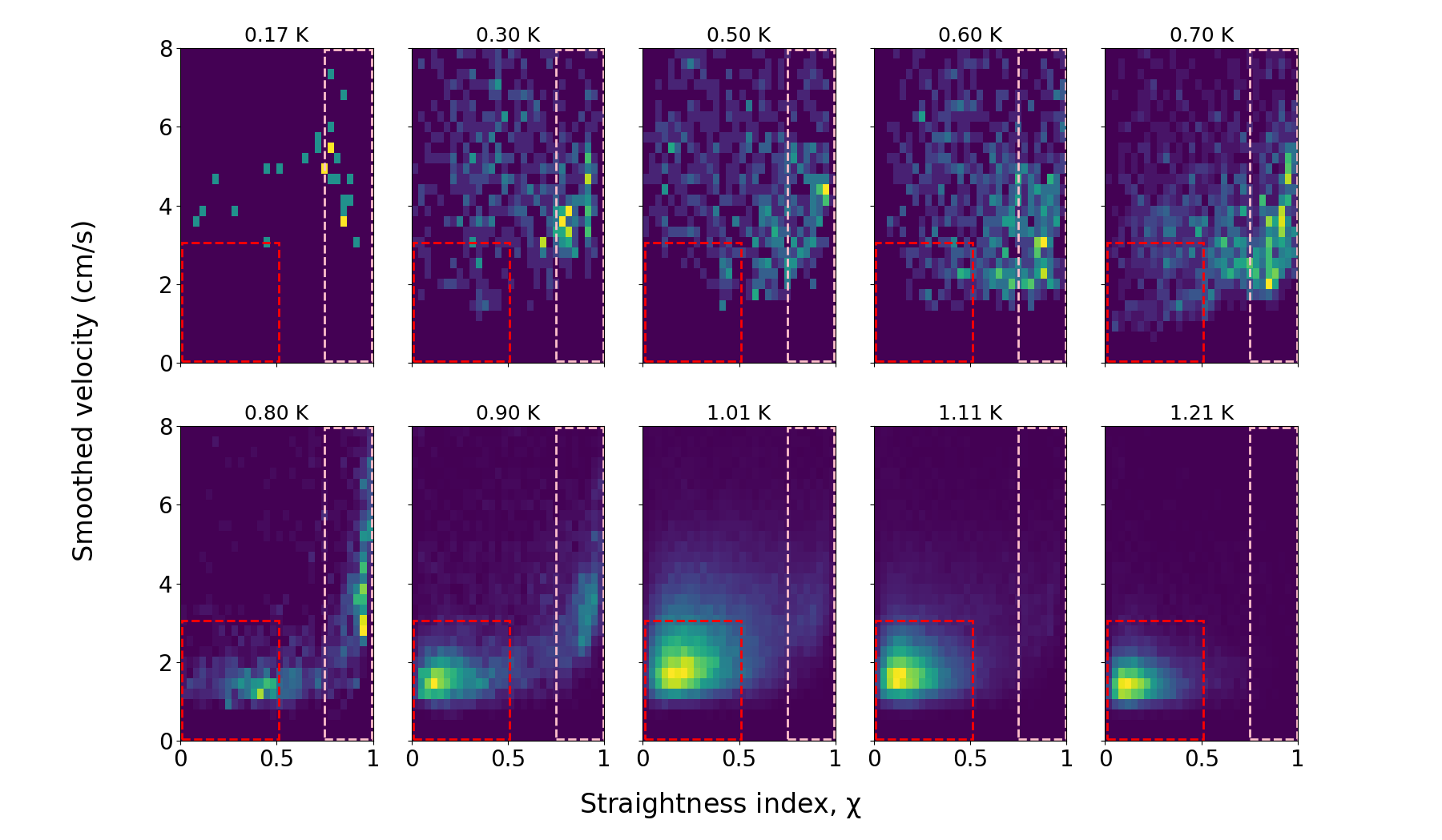}
    \caption{Particle binding parameter space shown for small bursts of 1--5 \textmu m particles. Smoothed velocity defined as $v = \delta r/\tau$ with $\tau = 10.1$ ms.  Heat map indicates linearly scaled occupation.  Dashed line boxes show the bound region (red) and free region (pink).}    
    \label{fig:binding space 1-5 1}
\end{figure}
\begin{figure}[h]
    \centering    \includegraphics[width=\linewidth]{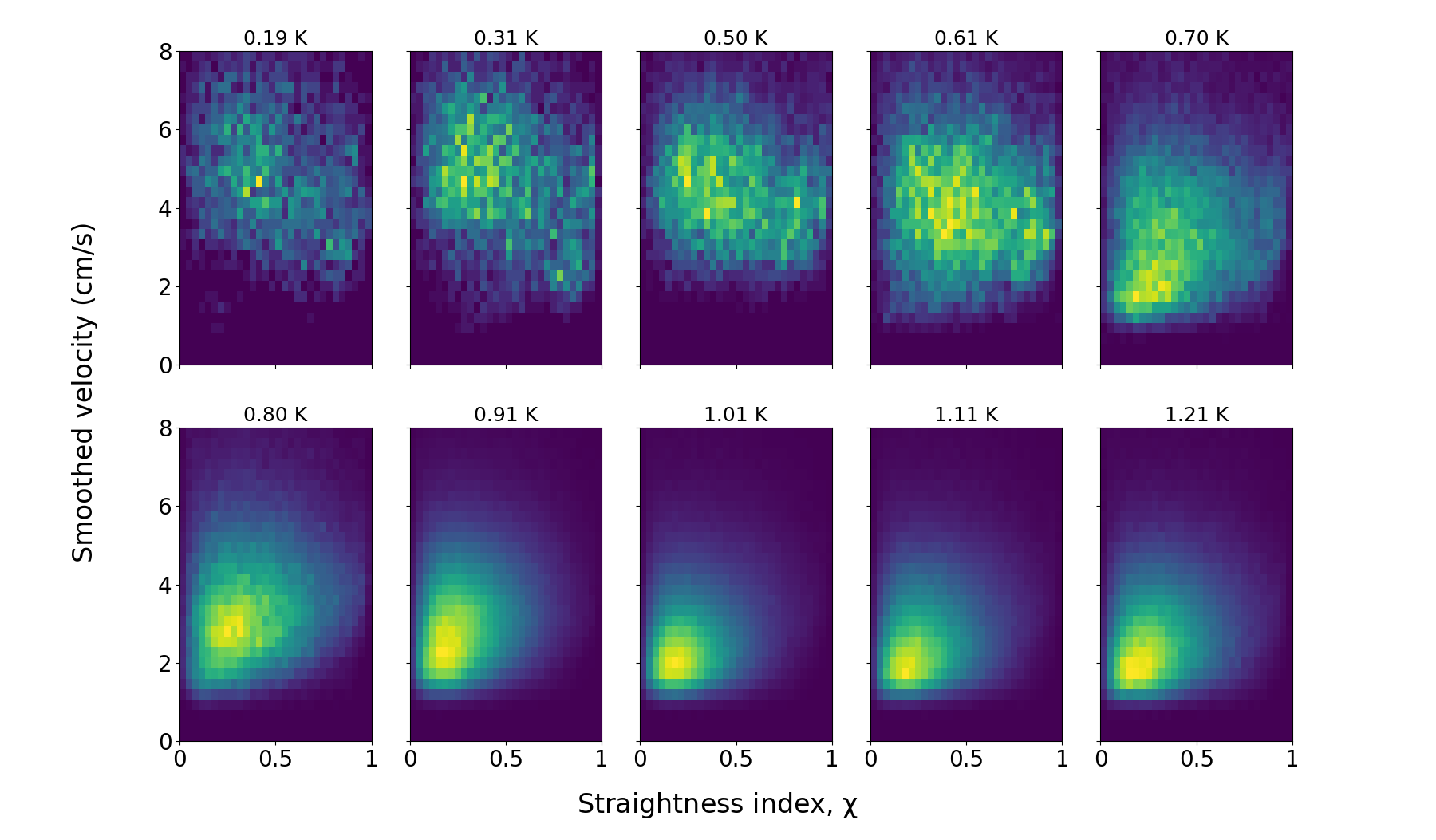}
    \caption{Particle binding parameter space shown for large bursts of 1--5 \textmu m particles. Smoothed velocity defined as $v = \delta r/\tau$ with $\tau = 10.1$ ms.  Heat map indicates linearly scaled occupation.}    
    \label{fig:binding space 1-5 2}
\end{figure}

We observed in Sec.\,\ref{Individual particle motions} that individual particles moved with erratic trajectories and speculated that such particles were bound on vortex lines undergoing a rapid series of reconnections with neighboring lines.  As these erratic particles have a slower drift velocity than those which appear to freely flow with the normal component.  We define the straightness index $\chi$ of the particle trajectory as
\begin{equation}
    \chi = \frac{R}{S},
    \label{straightness index}
\end{equation}
where $R$ is the magnitude of the particle's displacement and $S$ is the total point-to-point distance traveled between frames.  Defining a ``particle binding parameter space'', comprised of the smoothed (drift) velocity with $\tau = 10.1$ ms and the straightness index, two classes of particles emerge.  

In the case of 6 \textmu m particles as shown in Fig.\,\ref{fig:binding space 6 um},  the majority of entries occupy the $\chi > 0.75$ quadrant, indicating most particles follow smooth trajectories at all temperatures.  For 1--5 \textmu m particles the population of the space is more strongly temperature dependent.  Data for small bursts of particles are shown in Fig.\,\ref{fig:binding space 1-5 1}, and for large bursts in Fig.\,\ref{fig:binding space 1-5 2}.  At low temperatures, particle detections populate the space fairly randomly.  Note that at very low temperatures, there are few tracks with sufficiently long detectable paths to measure velocities on 10 frame time scales, particularly in the case of small bursts of particles.  At low temperatures, the detections shown in Fig.\,\ref{fig:binding space 1-5 1} mostly to aggregate in the $\chi > 0.75$ quadrant as most particles move freely with smooth trajectories. For small bursts of particles at $T > 0.8$ K, the particles strongly interact with vortices in greater numbers and begin to accumulate in what will be referred to as the bound region ($v < 3$\,cm\,s$^{-1}$, $\chi < 0.5$), eventually saturating in this area at $T > 1$ K.  In the case of large particle bursts shown in Fig.\,\ref{fig:binding space 1-5 2}, this bias toward the lower left region occurs from lower temperatures of $T > 0.6$\,K.  This is potentially due to the release of large bursts of particles (which requires higher transducer powers for longer durations) resulting in a more dense turbulent tangle of vortex lines developing in the cell.

%% file: Time_dependent_behaviour.tex
\section{Time dependent behavior} \label{Time dependent behaviour}
The impulsive injection of particles generates turbulence in the cell and results in chaotic particle motion at early times.  After $\sim 0.1$\,s the particles relax into a steady state dominated by a downward flow with a horizontal drift.  After the initial chaotic period, two primary time dependent effects are observed in the velocity distributions of particles and the number in the bound region of the binding parameter space shown in Figs.\,\ref{fig:binding space 1-5 1} and \ref{fig:binding space 1-5 2}.  
\begin{figure}[h]
    \centering    \includegraphics[width=\linewidth]{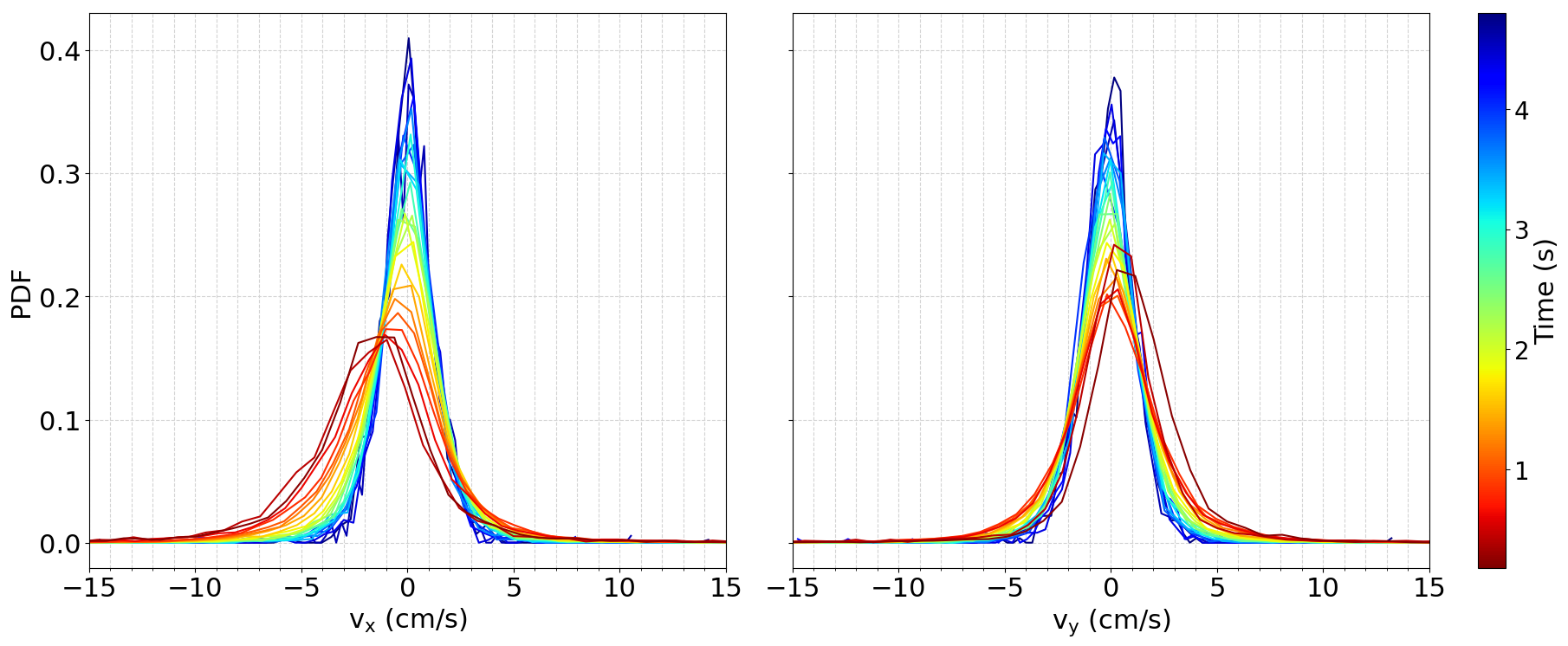}
    \caption{Evolution of velocity PDFs for 1--5 \textmu m particles in small bursts at 1 K.}    
    \label{fig:velocity time slices}
\end{figure}

\begin{figure}[h]
    \centering    \includegraphics[width=\linewidth]{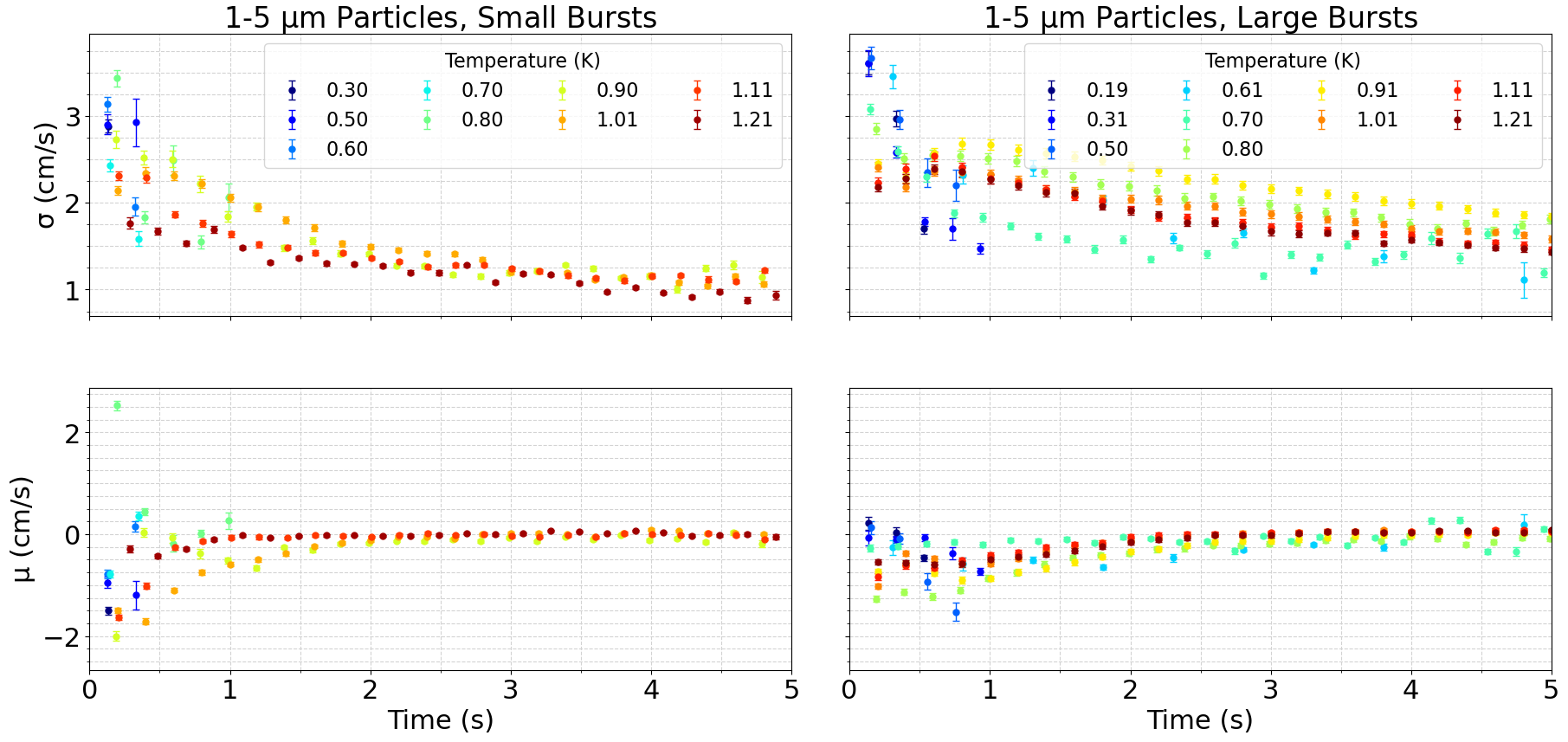}
    \caption{Gaussian fitting parameters for evolving horizontal ($x$) velocity PDFs for 1--5 \textmu m particles.}   
    \label{fig:fitted vx time slices}
\end{figure}

\begin{figure}[h]
    \centering    \includegraphics[width=\linewidth]{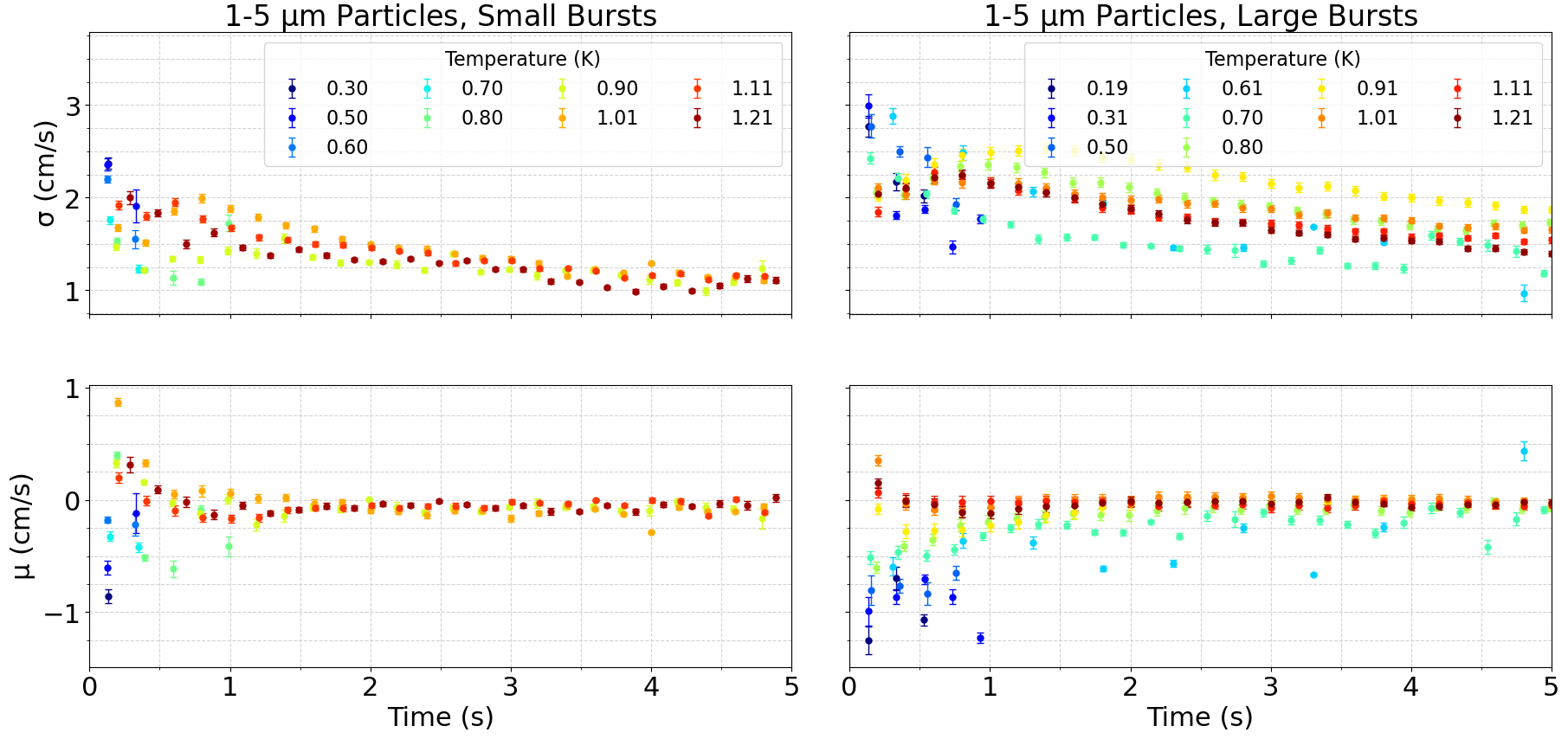}
    \caption{Gaussian fitting parameters for evolving vertical ($y$) velocity PDFs for 1--5 \textmu m particles.}   
    \label{fig:fitted vy time slices}
\end{figure}

By parsing the data into 0.2 second increments, beginning after the end of the injection period, the time dependence of the velocity distributions and the fraction of particles bound to vortices in the tangle may be observed.  Fig. \ref{fig:velocity time slices} shows the evolution of the velocity distribution of 1--5 \textmu m particles, released in small bursts at 1 K.  At late times the distributions have a narrow width and low mean horizontal velocity compared to early times.  Comparing to the raw footage (see supplementary material), this appears to be a reflection of the fact that at early times, the majority of particles are moving freely.  At later times, the free particles have almost entirely disappeared from the field of view and only particles suspended in the middle of the cell remain.  A clearer picture is obtained by fitting the central Gaussian part of the velocity PDFs and plotting the mean, $\mu$, and width, $\sigma$, as functions of time as shown in Figs. \ref{fig:fitted vx time slices} and \ref{fig:fitted vy time slices}.  

Clearly visible in the time dependence of the Gaussian parameters is the strong temperature dependence on the longevity of visible particles.  This is due to the effect of temperature on the drag on free particles affecting the particle drift velocities, preventing them from escaping the field of view.  They remain visible for the full video duration at temperatures above 0.7\,K and for only $\sim 1$ s at the lowest temperatures.  

\begin{figure}[h]
    \centering    \includegraphics[width=\linewidth]{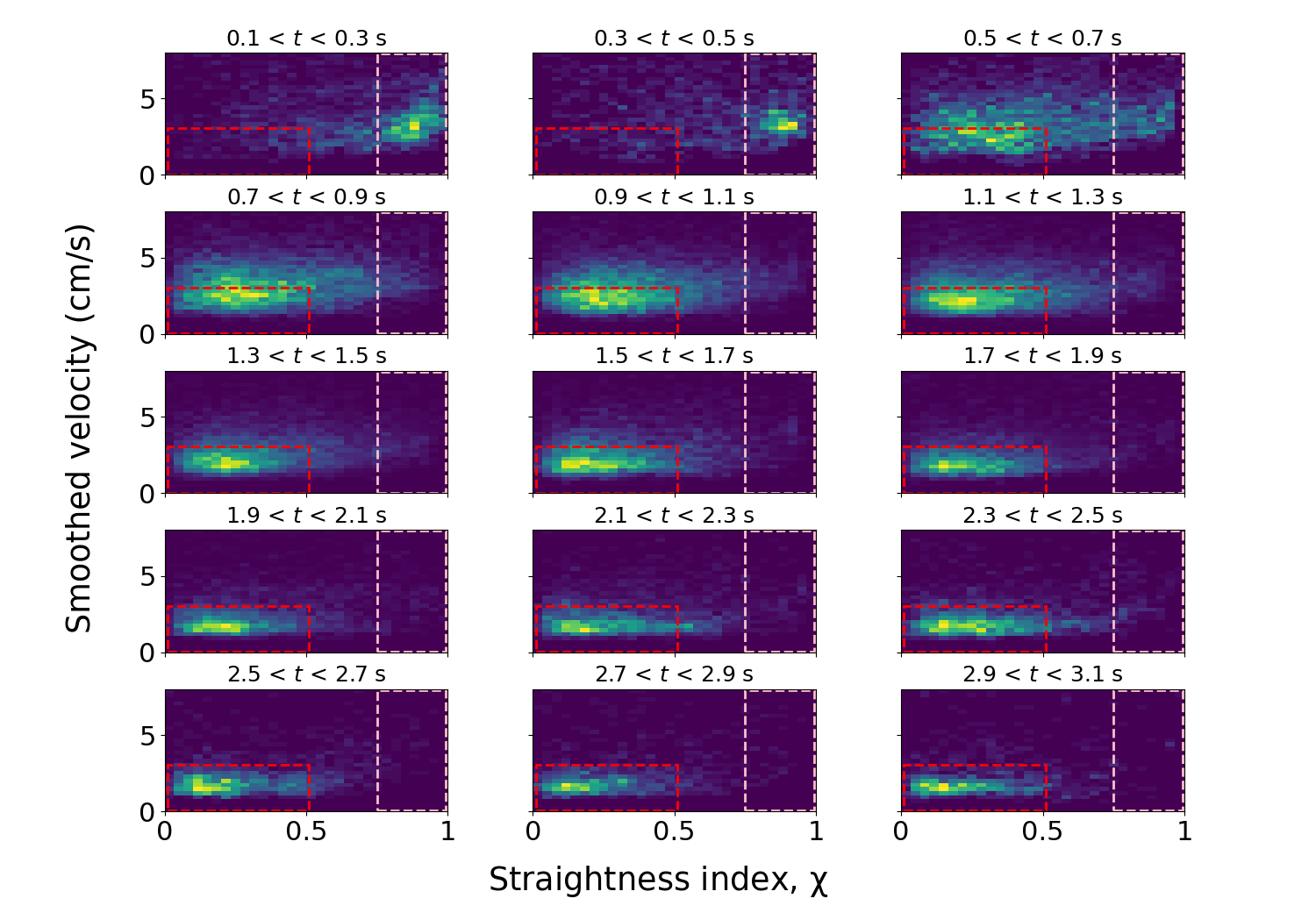}
    \caption{Evolution of binding space occupation for 1--5 \textmu m particles in small bursts at 1 K over first 3 seconds of after release of burst. Dashed line boxes show the bound region (red) and free region (pink).}    
    \label{fig:binding space evolution example}
\end{figure}

To test the hypothesis of the vortex tangle capturing large numbers of particles at late times, we use the binding space defined in Sec. \ref{Identifying particles bound to vortex lines} on successive 0.2\,s windows.  Fig. \ref{fig:binding space evolution example} shows the occupation of the binding space over the initial 3\,s after the chaotic injection period for small bursts of particles at 1\,K.  A clear shift in behavior is observed as particles migrate from the free quadrant ($\chi > 0.75$) to the bound region ($v(\tau = 10.1$\,ms$) < 3$\,cm\,s$^{-1}$, $\chi < 0.5$) over time. 

We investigated the temporal evolution of the observable particles following a large injection burst. The total number of particles in view at time $t$ is denoted by $N_t(t)$, while the number of particles in the bound region is denoted $N_b(t)$.  At temperatures below 0.8\,K, $N_t$ was decreasing too rapidly, and $N_b$ was typically near zero, thus preventing any quantitative analysis.  At temperatures above 1.0\,K, large classical eddies were frequently observed due to the turbulent injection process, and hence $N_b(t)$ was artificially inflated due to the circular motion of particles trapped in the jet but not bound on vortices.  We therefore focus on the temperature range 0.8--1.0\,K. 

\begin{figure}[h]
    \includegraphics[width=0.5\linewidth]{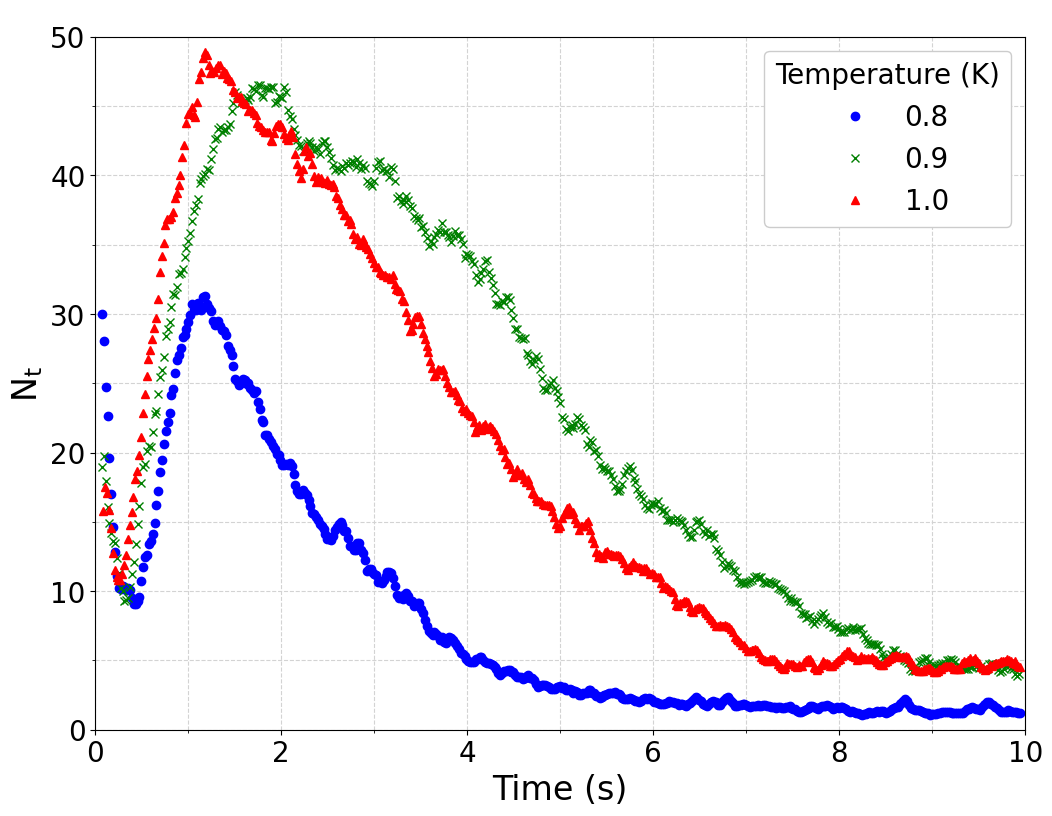}\includegraphics[width=0.5\linewidth]{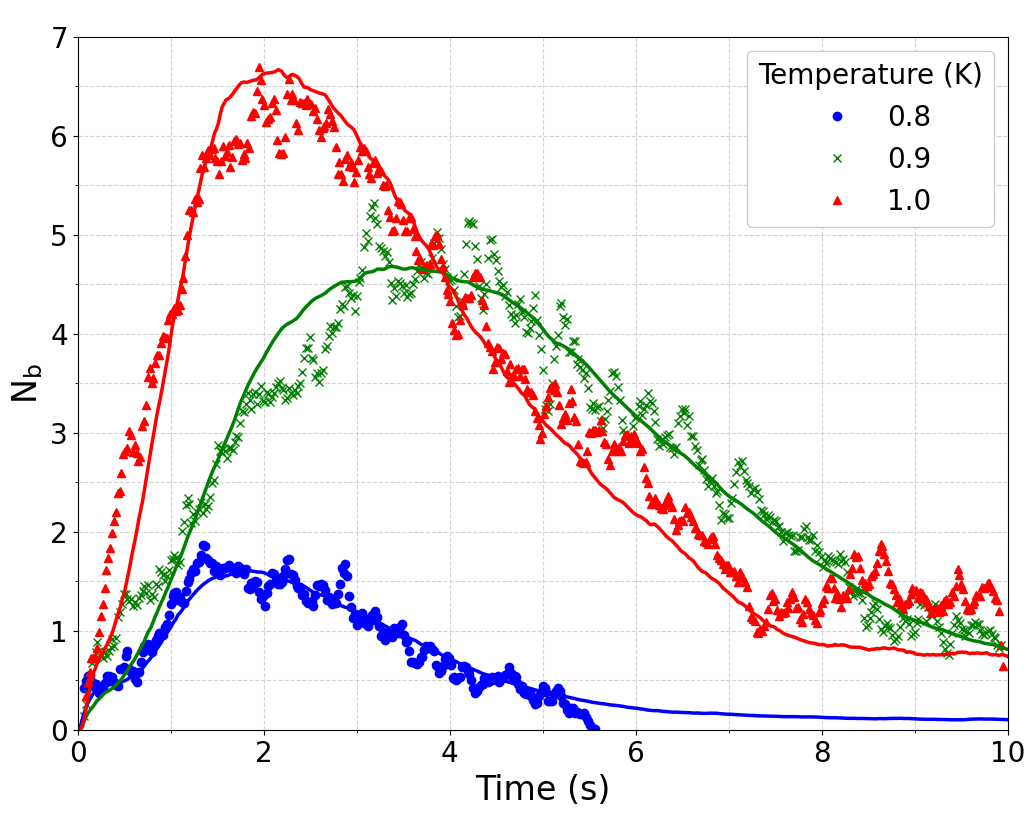}
    \caption{Top: Time trace for the total number of 1--5 \textmu m particles in large bursts at three example temperatures.  Bottom: Time trace of the number of bound particles (data points), compared to the results of numerical simulations (solid line). Experimental noise was reduced by averaging $N_b(t)$ over a 0.1\,s rolling window. For $T=0.8,$ 0.9, and 1\,K results, the values of $\omega_{trap}$ used were 0.1, 0.1 and 0.3\,s$^{-1}$, and those of $\omega_{free}$ were 1.4, 0.8, and 1.9\,s$^{-1}$ respectively.}   
    \label{fig:Numerical modelling of bound particles}
\end{figure}

The total number $N_t(t)$, shown in Fig. \ref{fig:Numerical modelling of bound particles}, often revealed a complex, non-monotonic transient behavior of duration
$\sim1$\,s, which we associate with the dynamics of the turbulent injection jet which brings them into the
region of observation. After passing through the peak value of $\sim 50 \pm 20$ particles, the transient changes to a
monotonic decay. On the other hand, the number of bound particles $N_b(t)$ always grew from $N_b(0) = 0$
until reaching a peak value of $N_b(t_{peak}) \sim 4 \pm 2$ particles, before switching to a nearly-exponential decay at later times.

To gain insight into the processes involved, we assumed that the vortex tangle is relatively slow
and that the probability per unit time that a free particle gets trapped by the tangle is $\omega_{trap}$, while the
rate of the trapped particles breaking free is $\omega_{free}$. Hence, $\dot{N_b} = \omega_{trap} (N_t - N_b) - \omega_{free} N_b$, where $\dot{N_b} = dN_b/dt$.  With assumption $N_b \ll N_t$, this becomes 
\begin{equation}
    \dot{N_b} = \omega_{trap} N_t - \omega_{free} N_b.
    \label{TrappingModel}
\end{equation}
Using the experimental value of $N_t(t)$ and the initial condition $N_b(0) = 0$, Eqn. \ref{TrappingModel} was solved numerically.  In the solutions, $N_b(t)$ go through a peak followed by an exponential decay at late times – much like the experimental $N_b(t)$. Examples are shown in Fig. \ref{fig:Numerical modelling of bound particles}.  By varying the parameters $\omega_{trap}$ and $\omega_{free}$, one can often reproduce the conspicuous features of the experimental $N_b(t)$: initial rate of increase, peak’s position and magnitude, late-time decay.  The values of $\omega_{trap}$ and $\omega_{free}$ used in the curves are shown in Fig. \ref{fig:Numerical modelling of bound particles}.
For some temperatures all these features could be captured by a suitable combination of $\omega_{trap}$ and $\omega_{free}$ while others had only limited success. Yet, the nature of the solution is such that it is not possible to accurately pinpoint optimal values of $\omega_{trap}$ and $\omega_{free}$ independently, especially because of the considerable noise in the experimental data. We would conclude that our naive model has the potential to capture the important aspects of the observed dynamics, but would require further refinement. Likewise, the quality of the experimental data would need to be improved for any
quantitative analysis to be performed.

%% file: Conclusions.tex
\section{Conclusion}

We have developed a system for imaging small polymer tracer particles in turbulent superfluid \4 at temperatures as low as 0.14\,K.  By studying the statistical behavior of particles through their velocity distributions and the shapes of their trajectories, we have investigated their interactions with both the normal component and quantized vortices in the superfluid component. Certain features, known from the studies of counterflow turbulence at temperatures above 1.6\,K have been confirmed, such as the bimodal distribution of the particle velocities in the direction of counterflow \cite{bewley2006, mastracci2019characterizing, vsvanvcara2021ubiquity} and the sensitivity of power-law tails in the velocity distributions to the length scale probed \cite{paoletti2008PowerLaw, mantia2014a, mantia2014b}. 

We have identified a number of novel features in the distributions observed some of which have actually been seen in numerical simulations of vortex tangles.  These include exponential distribution of particles velocities in alignment with the turbulent velocities simulated using the vortex filament model \cite{baggaley2014acceleration, galantucci2025quantum}, a strong temperature dependence in the trapping efficiency of particles \cite{poole2005motion} and an upper limit of 0.6\,K on the minimum temperature at which particles can become stably bound to vortex lines in statistically significant numbers.

In addition to the temperature dependence of particle binding, we have observed the time dependent populations of free and bound particles, and shown that in the temperature range 0.8--1.0\,K the rates of trapping and de-trapping can be simulated using an analytical model.

\begin{acknowledgments}
The financial support came from EPSRC through grant No. EP/P025625/1. W. Guo acknowledges support from the Gordon and Betty Moore Foundation through grant DOI 10.37807/gbmf11567 and the National High Magnetic Field Laboratory at Florida State University, which is supported by the National Science Foundation Cooperative Agreement No. DMR-2128556 and the State of Florida. 
\end{acknowledgments}


\section*{Data Availability}
The data that support the findings of this study are available from the corresponding author upon reasonable request.